\documentclass[twocolumn]{aastex6}
\usepackage{graphicx}
\usepackage[flushleft]{threeparttable}
\shorttitle{ASPECS: CO LFs and $\rho$(H$_2$) vs.~$z$}
\shortauthors{Decarli et al.}

\def\Msun{M$_\odot$}

\def\Ci{[C\,{\sc i}]}

\def\Cii{[C\,{\sc ii}]}
\def\Ci{[C\,{\sc i}]}

\def\kms{km\,s$^{-1}$}

\def\Kkmspc{K~km\,s$^{-1}$\,pc$^2$}
\def\lsim{\mathrel{\rlap{\lower 3pt \hbox{$\sim$}} \raise 2.0pt \hbox{$<$}}}
\def\gsim{\mathrel{\rlap{\lower 3pt \hbox{$\sim$}} \raise 2.0pt \hbox{$>$}}}

\begin{document}

\title{The ALMA Spectroscopic Survey in the HUDF: CO luminosity functions and the molecular gas content of galaxies through cosmic history}

\author{Roberto Decarli\altaffilmark{1},
Fabian Walter\altaffilmark{2,3},
Jorge G\'onzalez-L\'opez\altaffilmark{4,5}, % 0000-0003-3926-1411
Manuel Aravena\altaffilmark{4},
Leindert Boogaard\altaffilmark{6},
Chris Carilli\altaffilmark{3,7},
Pierre Cox\altaffilmark{8},
Emanuele Daddi\altaffilmark{9},
Gerg\"{o} Popping\altaffilmark{2},
Dominik Riechers\altaffilmark{10,2},
Bade Uzgil\altaffilmark{3,2},
Axel Weiss\altaffilmark{11},
Roberto J. Assef\altaffilmark{4},
Roland Bacon\altaffilmark{12},
Franz Erik Bauer\altaffilmark{5,13,14},
Frank Bertoldi\altaffilmark{15},
Rychard Bouwens\altaffilmark{5},
Thierry Contini\altaffilmark{16},
Paulo C.~Cortes\altaffilmark{17,18},
Elisabete da Cunha\altaffilmark{19},
Tanio D\'iaz-Santos\altaffilmark{4},
David Elbaz\altaffilmark{8},
Hanae Inami\altaffilmark{11,20},
Jacqueline Hodge\altaffilmark{5},
Rob Ivison\altaffilmark{21,22},
Olivier Le F\`evre\altaffilmark{23},
Benjamin Magnelli\altaffilmark{15},
Mladen Novak\altaffilmark{2},
Pascal Oesch\altaffilmark{24},
Hans--Walter Rix\altaffilmark{2},
Mark T. Sargent\altaffilmark{25},
Ian R.~Smail\altaffilmark{26},
A. Mark Swinbank\altaffilmark{27},
Rachel S. Somerville\altaffilmark{27,28},
Paul van der Werf\altaffilmark{5},
Jeff Wagg\altaffilmark{29},
Lutz Wisotzki\altaffilmark{30}
} % [0000-0002-2662-8803]
\altaffiltext{1}{INAF---Osservatorio di Astrofisica e Scienza dello Spazio, via Gobetti 93/3, I-40129, Bologna, Italy. E-mail: {\sf roberto.decarli@inaf.it}}
\altaffiltext{2}{Max Planck Institut f\"ur Astronomie, K\"onigstuhl 17, 69117 Heidelberg, Germany}
\altaffiltext{3}{National Radio Astronomy Observatory, Pete V. Domenici Array Science Center, P.O. Box O, Socorro, NM 87801, USA}
\altaffiltext{4}{N\'{u}cleo de Astronom\'{\i}a, Facultad de Ingenier\'{\i}a y Ciencias, Universidad Diego Portales, Av. Ej\'{e}rcito 441, Santiago, Chile}
\altaffiltext{5}{Instituto de Astrof\'{\i}sica, Facultad de F\'{\i}sica, Pontificia Universidad Cat\'olica de Chile Av. Vicu\~na Mackenna 4860, 782-0436 Macul, Santiago, Chile}
\altaffiltext{6}{Leiden Observatory, Leiden University, PO Box 9513, NL-2300 RA Leiden, The Netherlands}
\altaffiltext{7}{Battcock Centre for Experimental Astrophysics, Cavendish Laboratory, Cambridge CB3 0HE, UK}
\altaffiltext{8}{Institut d'Astrophysique de Paris, Sorbonne Universit\'{e}, CNRS, UMR 7095, 98 bis bd Arago, 7014 Paris, France}
\altaffiltext{9}{Laboratoire AIM, CEA/DSM-CNRS-Universite Paris Diderot, Irfu/Service d'Astrophysique, CEA Saclay, Orme des Merisiers, 91191 Gif-sur-Yvette cedex, France}
\altaffiltext{10}{Cornell University, 220 Space Sciences Building, Ithaca, NY 14853, USA}
\altaffiltext{11}{Max-Planck-Institut f\"ur Radioastronomie, Auf dem H\"ugel 69, 53121 Bonn, Germany}
\altaffiltext{12}{Univ. Lyon 1, ENS de Lyon, CNRS, Centre de Recherche Astrophysique de Lyon (CRAL) UMR5574, 69230 Saint-Genis-Laval, France}
\altaffiltext{13}{Millennium Institute of Astrophysics (MAS), Nuncio Monse{\~{n}}or S{\'{o}}tero Sanz 100, Providencia, Santiago, Chile}
\altaffiltext{14}{Space Science Institute, 4750 Walnut Street, Suite 205, Boulder, CO 80301, USA}
\altaffiltext{15}{Argelander-Institut f\"ur Astronomie, Universit\"at Bonn, Auf dem H\"ugel 71, 53121 Bonn, Germany}
\altaffiltext{16}{Institut de Recherche en Astrophysique et Plan\'{e}tologie (IRAP), Universit\'{e} de Toulouse, CNRS, UPS, 31400 Toulouse, France}
\altaffiltext{17}{Joint ALMA Observatory - ESO, Av. Alonso de C\'ordova, 3104, Santiago, Chile}
\altaffiltext{18}{National Radio Astronomy Observatory, 520 Edgemont Rd, Charlottesville, VA, 22903, USA}
\altaffiltext{19}{Research School of Astronomy and Astrophysics, Australian National University, Canberra, ACT 2611, Australia}
\altaffiltext{20}{Hiroshima Astrophysical Science Center, Hiroshima University, 1-3-1 Kagamiyama, Higashi-Hiroshima, Hiroshima, 739-8526, Japan}
\altaffiltext{21}{European Southern Observatory, Karl-Schwarzschild-Strasse 2, 85748, Garching, Germany}
\altaffiltext{22}{Institute for Astronomy, University of Edinburgh, Royal Observatory, Blackford Hill, Edinburgh EH9 3HJ}
\altaffiltext{23}{Aix Marseille Universit\'e, CNRS, LAM (Laboratoire d'Astrophysique de Marseille), UMR 7326, F-13388 Marseille, France}
\altaffiltext{24}{Department of Astronomy, University of Geneva, Ch. des Maillettes 51, 1290 Versoix, Switzerland}
\altaffiltext{25}{Astronomy Centre, Department of Physics and Astronomy, University of Sussex, Brighton, BN1 9QH, UK}
\altaffiltext{26}{Centre for Extragalactic Astronomy, Department of Physics, Durham University, South Road, Durham, DH1 3LE, UK}
\altaffiltext{27}{Department of Physics and Astronomy, Rutgers, The State University of New Jersey, 136 Frelinghuysen Rd, Piscataway, NJ 08854, USA}
\altaffiltext{28}{Center for Computational Astrophysics, Flatiron Institute, 162 5th Ave, New York, NY 10010, USA}
\altaffiltext{29}{SKA Organization, Lower Withington Macclesfield, Cheshire SK11 9DL, UK}
\altaffiltext{30}{Leibniz-Institut f\"{u}r Astrophysik Potsdam, An der Sternwarte 16, 14482 Potsdam, Germany}

\begin{abstract}
We use the results from the ALMA large program ASPECS, the spectroscopic survey in the Hubble Ultra Deep Field (HUDF), to constrain CO luminosity functions of galaxies and the resulting redshift evolution of $\rho$(H$_2$). The broad frequency range covered enables us to identify CO emission lines of different rotational transitions in the HUDF at $z>1$. We find strong evidence that the CO luminosity function evolves with redshift, with the knee of the CO luminosity function decreasing in luminosity by an order of magnitude from $\sim$2 to the local universe. Based on Schechter fits, we estimate that our observations recover the majority (up to $\sim$90\%, depending on the assumptions on the faint end) of the total cosmic CO luminosity at $z$=1.0--3.1. After correcting for CO excitation, and adopting a Galactic CO--to--H$_2$ conversion factor, we constrain the evolution of the cosmic molecular gas density  $\rho$(H$_2$): this cosmic gas density peaks at $z\sim1.5$ and drops by factor of $6.5_{-1.4}^{+1.8}$ to the value measured locally. The observed evolution in $\rho$(H$_2$) therefore closely matches the evolution of the cosmic star formation rate density $\rho_{\rm SFR}$. We verify the robustness of our result with respect to assumptions on source inclusion and/or CO excitation. As the cosmic star formation history can be expressed as the product of the star formation efficiency and the cosmic density of molecular gas, the similar evolution of $\rho$(H$_2$) and $\rho_{\rm SFR}$ leaves only little room for a significant evolution of the average star formation efficiency in galaxies since $z\sim 3$ (85\% of cosmic history). 
\end{abstract} 

\section{Introduction}

The molecular phase of the interstellar medium (ISM) is the birthplace of stars, and therefore it plays a central role in the evolution of galaxies \citep[see reviews in][]{kennicutt12,carilli13,bolatto13}. The cosmic history of star formation \citep[see, e.g.,][]{madau14}, i.e., the mass of stars formed per unit time in a cosmological volume (or cosmic star formation rate density, $\rho_{\rm SFR}$) throughout cosmic time, increased from early cosmic epochs up to a peak at $z$=1--3, and then declined by a factor $\sim$8 until the present day. This could be explained by a larger supply of molecular gas (the fuel for star formation) in high--$z$ galaxies; by physical properties of the gas, that could more efficiently form stars; or by a combination of both. The characterization of the content and properties of the molecular ISM in galaxies at different cosmic epochs is therefore fundamental to our understanding of galaxy formation and evolution.

The H$_2$ molecule, the main constituent of molecular gas, is a poor radiator: it lacks rotational transitions, and the energy levels of vibrational lines are populated significantly only at relatively high temperatures ($T_{\rm ex}>500$\,K) that are not typical of the cold, star--forming ISM \citep{omont07}. On the other hand, the carbon monoxide molecule, $^{12}$CO (hereafter, CO) is the second most abundant molecule in the universe. Thanks to its bright rotational transitions, it has been detected even at the highest redshifts \citep[$z\sim 7$; e.g.,][]{riechers13,venemans17a,strandet17,marrone18}. Redshifted CO lines are observed in the radio and millimeter (mm) transparent windows of the atmosphere, thus becoming accessible to facilities such as the Jansky Very Large Array (JVLA), the IRAM NOrthern Expanded Millimeter Array (NOEMA), and the Atacama Large Millimeter Array (ALMA). CO is therefore the preferred observational probe of the molecular gas content in galaxies at high redshift.

To date, more than 250 galaxies have been detected in CO at $z>1$, the majority of which are quasar host galaxies or sub-mm galaxies (see, \citealt{carilli13} or a review); gravitationally--lensed galaxies (e.g., \citealt{riechers10}, \citealt{harris12}, \citealt{dessaugeszavadsky15}, ; \citealt{aravena16c}; \citealt{dessaugeszavadsky17}, \citealt{gonzalezlopez17}); and (proto-)clusters of galaxies (e.g., \citealt{aravena12}, \citealt{chapman15}, \citealt{seko16}, \citealt{rudnick17}, \citealt{hayatsu17}, \citealt{lee17}, \citealt{hayashi18}, \citealt{miller18}, \citealt{oteo18}). The remainder are galaxies selected based on their stellar mass ($M_{\rm *}$), star formation rate (SFR), and/or optical/near-infrared colors \citep[e.g.,][]{daddi10a, daddi10b, tacconi10, tacconi13, tacconi18, genzel10, genzel11, genzel15}. These studies were instrumental in shaping our understanding of the interplay between molecular gas reservoirs and star formation in massive $z>1$ galaxies on and above the `main sequence' of star-forming galaxies \citep{noeske07,elbaz11}. E.g., these galaxies are found to have high molecular gas fractions $M_{\rm H2}$/$M_{\rm *}$ compared to galaxies in the local universe. The depletion time, $t_{\rm dep}$=$M_{\rm H2}$/SFR, i.e., the time required to consume the entire molecular gas content of a galaxy at the present rate of star formation, is shorter in starburst galaxies than in galaxies on the main sequence \citep[see, e.g.,][]{silverman15,silverman18,schinnerer16,scoville17,tacconi18}. However, by nature these targeted studies are potentially biased towards specific types of galaxies (e.g., massive, star-forming galaxies), and consequently might fail to capture the full diversity of gas-rich galaxies in the universe.

Spectral line scans provide a complementary approach. These are interferometric observations over wide frequency ranges, targeting `blank' regions of the sky. Gas, traced mainly via CO lines, is searched for at any position and frequency, without pre-selection based on other wavelengths. This provides us with a flux--limited census of the gas content in well--defined cosmological volumes. The first molecular scan reaching sufficient depth to detect MS galaxies targeted a $\sim$1\,arcmin$^{2}$ region in the {\em Hubble} Deep Field North \citep[HDF-N;][]{williams96} using the IRAM Plateau de Bure Interferometer (PdBI; see \citealt{decarli14}). The scan resulted in the first redshift measurement for the archetypal sub-mm galaxy HDF\,850.1 ($z$=$5.183$, see \citealt{walter12}), and in the discovery of massive ($>10^{10}$\,\Msun{}) gaseous reservoirs associated with galaxies at $z\sim 2$, including one with no obvious optical/NIR counterpart \citep{decarli14}. These observations enabled the first, admittedly loose constraints on the CO luminosity functions (LFs) and on the cosmic density of molecular gas in galaxies, $\rho$(H$_2$), as a function of redshift \citep{walter14}. The HDF-N was also part of a second large observing campaign using the JVLA, the COLDz project. This effort ($>300$\,hr of observations) targeted a $\sim$48\,arcmin$^2$ area in the GOODS-North footprint \citep{giavalisco04}, and a $\sim8$\,arcmin$^2$ region in COSMOS \citep{scoville07}, sampling the frequency range 30--38\,GHz \citep{lentati15,pavesi18}. This exposed the CO(1-0) emission in galaxies at $z\approx2.0$--$2.8$ and the CO(2-1) emission at $z\approx4.9$--6.7. The unprecedentedly large area covered by COLDz resulted in the best constraints on the CO LFs at $z>2$ so far, especially at the bright end \citep{riechers19}. 

In ALMA Cycle 2, we scanned the 3\,mm and 1.2\,mm windows (84--115\,GHz and 212--272\,GHz, respectively) in a $\sim$1\,arcmin$^2$ region in the {\em Hubble} Ultra Deep Field \citep[HUDF;][]{beckwith06}. This pilot program, dubbed the ALMA Spectroscopic Survey in the HUDF \citep[ASPECS;][]{walter16}, pushed the constraints on the CO LFs at high redshift towards the expected knee of the CO LFs \citep{decarli16a}. By capitalizing on the combination of the 3\,mm and 1.2\,mm data, and on the unparalleled wealth of ancillary information available in the HUDF, \citet{decarli16b} were able to measure CO excitation in some of the observed sources, and to relate the CO emission to other properties of the observed galaxies at various wavelengths. Furthermore, the collapsed 1.2\,mm data cube resulted in the deepest dust continuum image ever obtained at these wavelengths ($\sigma$=13\,$\mu$Jy\,beam$^{-1}$), which allowed us to resolve $\sim 80$\% of the cosmic infrared background \citep{aravena16a}. The 1.2\,mm data were also exploited to perform a systematic search for \Cii{} emitters at $z$=6--8 \citep{aravena16b}, as well as to constrain the IRX--$\beta$ relation at high redshift \citep{bouwens16}. Finally, the ASPECS Pilot provided first direct measurements of the impact of foreground CO lines on measurements of the cosmic microwave background fluctuations, which is critical for intensity mapping experiments \citep{carilli16}.

\begin{figure*}
\begin{center}
\includegraphics[width=0.99\textwidth]{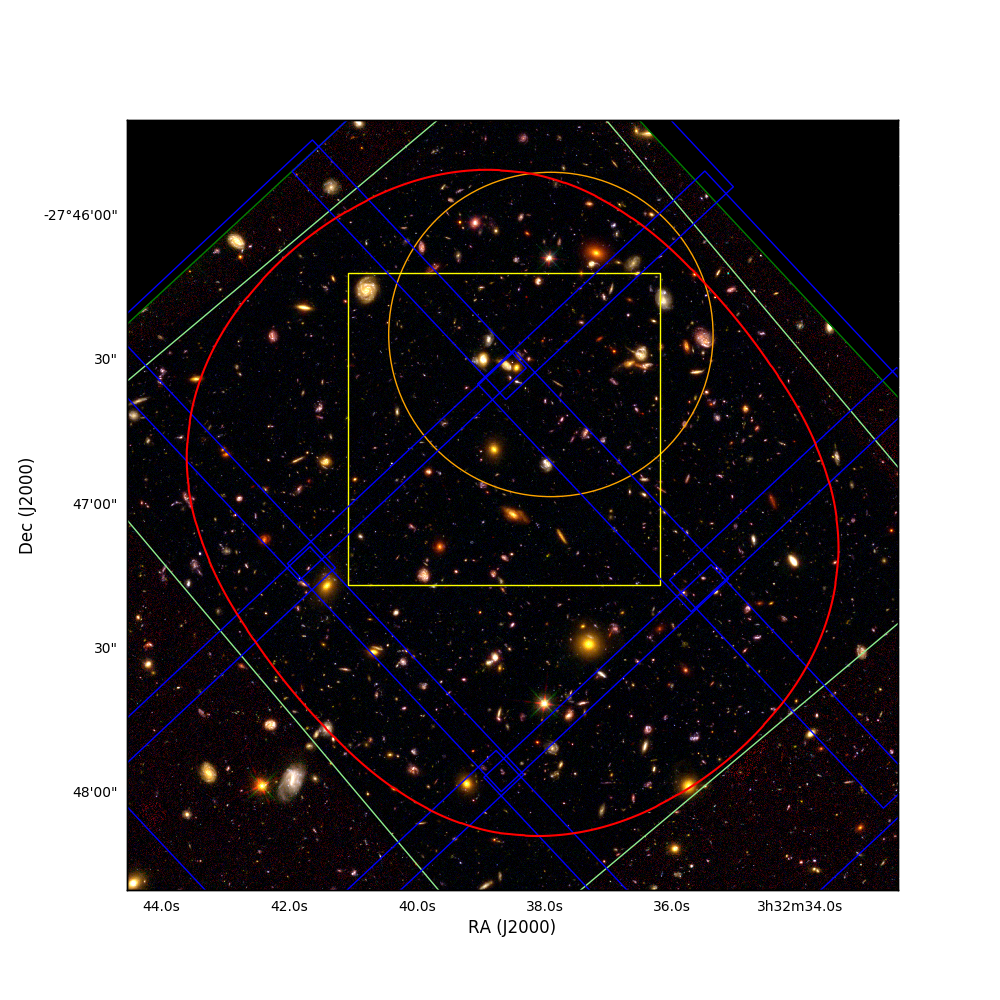}\\
\end{center}
\caption{{\em Hubble} RGB images (red: F105W filter, green: F770W filter, blue: F435W filter) of the {\em Hubble} Ultra Deep Field (dark green contour). For comparison, we plot the coverage of the {\em Hubble} eXtremely Deep Field \citep[XDF;][]{illingworth13,koekemoer13}, in light green; the pointings of the MUSE UDF survey \citep{bacon17}, in blue; the deep MUSE pointing \citep{bacon17} in yellow. The 50\% sensitivity contours of the ASPECS pilot \citep{walter16} and of the ASPECS LP 3\,mm survey are shown in orange and red, respectively (see also \citealt{gonzalezlopez19}). The area covered in our study encompasses $>$7000 catalogued galaxies, with hundreds of spectroscopic redshifts, and photometry in $>$30 bands.}
\label{fig_field}
\end{figure*}

The ASPECS Pilot program was limited by the small area surveyed. Here we present results from the ASPECS Large Program (ASPECS LP). The project replicates the survey strategy of the ASPECS Pilot, but on a larger mosaic that covers most of the {\em Hubble} eXtremely Deep Field (XDF), the region of the HUDF where the deepest near-infrared data are available (\citealt{illingworth13,koekemoer13}; see Fig.~\ref{fig_field}). Here we present and focus on the ASPECS LP 3\,mm data, which have been collected in ALMA Cycle 4. We discuss the survey strategy and observations, the data reduction, the ancillary dataset, and we use the CO detections from the 3\,mm data to measure the CO LFs in various redshift bins, and to infer the cosmic gas density $\rho$(H$_2$) as a function of redshift. In \citet{gonzalezlopez19} (hereafter, GL19), we present our search for line and continuum sources, and assess their reliability and completeness. \citet{aravena19} place the ASPECS LP 3\,mm results in the context of the main sequence narrative. \citet{boogaard19} capitalize on the sensitive VLT/MUSE Integral Field Spectroscopy of the field, in order to address how our CO detections relate with the properties of the galaxies as inferred from rest-frame optical/UV wavelengths. Finally, \citet{popping19} compare the ASPECS LP 3\,mm results to state-of-the-art predictions from cosmological simulations and semi-analytical models. 

The structure of this paper is as follows: In Sec.~\ref{sec_obs}, we present the survey strategy, the observations, and the data reduction. In Sec.~\ref{sec_ancillary} we summarize the ancillary information available for the galaxies in this field. In Sec.~\ref{sec_results} we present the main results of this study, and in Sec.~\ref{sec_discussion} we discuss our findings and compare them with similar works in the literature. Finally, in Sec.~\ref{sec_conclusions} we infer our conclusions.  

Throughout this paper we adopt a $\Lambda$CDM cosmological model with $H_0=70$ km\,s$^{-1}$\,Mpc$^{-1}$, $\Omega_{\rm m}=0.3$ and $\Omega_{\Lambda}=0.7$ \citep[consistent with the measurements by the][]{planck15}. Magnitudes are reported in the AB photometric system. For consistency with the majority of the literature on this field, in our analysis, we adopt a \citet{chabrier03} stellar initial mass function. % (IMF), although recent studies \citep[e.g.,][]{zhang18} suggest that top--heavy IMFs might be more appropriate for intense starbursts. Such IMFs would reduce the inferred SFRs, thus increasing the depletion time.

\section{Observations and data processing}\label{sec_obs}

\subsection{Survey design and observations}

The ASPECS LP survey consists of a 150\,hr program in ALMA Cycle 4 (Program ID: 2016.1.00324.L). ASPECS LP comprises two scans, at 3\,mm and 1.2\,mm. The 3\,mm survey presented here took 68\,hr of telescope time (including calibrations and overheads), and was executed between December 2--21, 2016 (ALMA Cycle 4). 

\begin{figure}
\begin{center}
\includegraphics[width=0.49\textwidth]{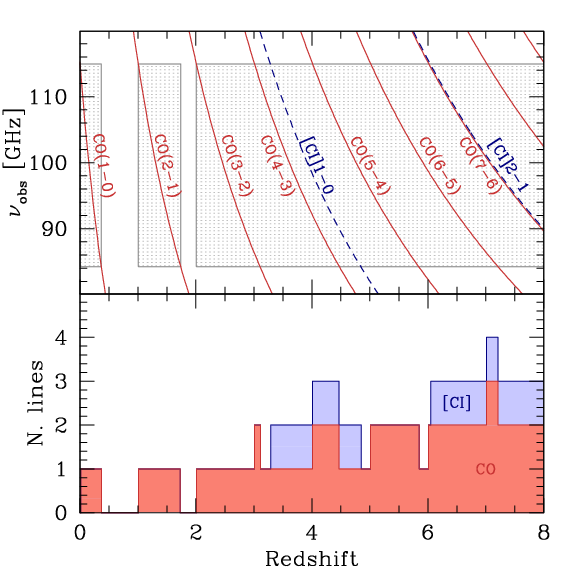}
\end{center}
\caption{{\em Top:} The observed frequency of various CO and \Ci{} transitions covered in our 3\,mm scan, as a function of redshift. The shaded area marks the parameter space sampled in our study. {\em Bottom:} Number of CO or \Ci{} transitions observable in our 3\,mm scan (exclusively based on frequency coverage), as a function of redshift. The frequency range encompassed in our study enables the detection of CO at $z\lsim0.37$, $1.0\lsim z\lsim 1.7$, and virtually at any $z\gsim 2.00$. Additionally, our scan covers 2 or more transitions at most redshifts above $z\sim3$. }
\label{fig_nlines_z}
\end{figure}

These observations comprised 17 pointings covering most of the XDF (\citealt{illingworth13,koekemoer13}; see Fig.~\ref{fig_field}). The pointings were arranged in a hexagonal pattern, distanced by $26.4''$ (the half--width of the primary beam of ALMA 12m antennas at the high--frequency end of ALMA band 3), thus ensuring Nyquist sampling and spatially--homogeneous noise in the mosaic. For reference, the central pointing is centered at Right Ascension = 03:32:38.5 and Declination = -27:47:00 (J2000.0). The total area covered at the center of the frequency scan ($\approx 99.5$\,GHz) with primary beam attenuation $<0.5$ is 4.6\,arcmin$^2$. The observing strategy capitalized on the fast slew of the ALMA antennas in order to fully cover the entire mosaic between each phase calibrator observation. The survey was executed with the array in a relatively compact (C40-3) configuration. Baselines ranged between 15 and 700\,m. The quasar J0334-4008 was observed as a flux, bandpass, and pointing calibrator, while the quasar J0342-3007 served as phase calibrator. The observations were performed in 5 different frequency settings, covering the frequency range 84--115\,GHz. This enables the observation of one or more CO lines over a wide range of redshifts (see Fig.~\ref{fig_nlines_z}). Lower and upper side bands of the frequency settings partially overlap in the central part of the frequency range (96--103\,GHz), thus yielding improved sensitivity at these frequencies (see also Fig.~3 in GL19).

\subsection{Data reduction, calibration, and imaging}

We processed the data using both the CASA pipeline for ALMA data \citep[v.~4.7.0;][]{mcmullin07} and our own procedures \citep[see, e.g.,][]{aravena16a}, which follow the general scheme of the official ALMA pipeline. Our independent inspection for data to be flagged allowed us to improve the depth of our scan in one of the frequency settings by up to 20\%. In all the other frequency settings, the final rms appears consistent with the one computed from the cube provided by the ALMA pipeline. As the cube created with our own procedures is at least as good (in terms of low noise) as the one from the pipeline, we will refer to the former in the remainder of the analysis.

\begin{figure}
\begin{center}
\includegraphics[width=0.99\columnwidth]{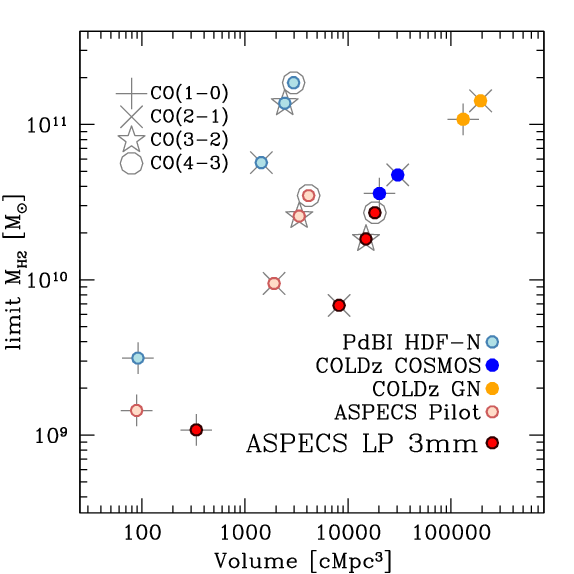}\\
\end{center}
\caption{Depth and volume coverage of the molecular scans performed so far: the PdBI scan \citep{decarli14,walter14}, the COLDz survey \citep{pavesi18,riechers19}, the ASPECS Pilot \citep{walter16,decarli16a}, and the ASPECS LP 3mm (this work). The H$_2$ mass limits are computed at 5-$\sigma$ in the case of line widths of 200\,\kms{}, assuming the CO SLED by \citet{daddi15} and a CO--to--H$_2$ conversion factor $\alpha_{\rm CO}$=$3.6$ \Msun{}\,(K\,\kms\,pc$^2$)$^{-1}$. Limits from various CO transitions are plotted. The complementarity of field coverage and depth in these campaigns is apparent.}
\label{fig_surveys}
\end{figure}

We imaged the 3\,mm cube with natural weighting using the task \textsf{tclean}. The resulting synthesized beam is $\approx 1.75''\times1.49''$ (PA=$91.5^\circ$) at the center of the observed frequency range. The lack of very bright sources in our cubes allows us to perform our analysis on the `dirty' cube, thus preserving the intrisic properties of the noise. The resulting cube is used for the line search, and in most of the following analysis. In addition, we image the dataset after applying a u,v taper of $3''$ (using the \textsf{uvtaper} parameter in \textsf{tclean}) cleaned to 2-$\sigma$. This yields a reconstructed beam of $\sim3.5''$. This latter cube is used only to extract the spectra of the sources identified in the search: thanks to the lower angular resolution, the spectra extracted in this way encapsulate all the emission of the sources, even in the case of sources that are spatially--resolved in the naturally--weighted imaging (see \citealt{aravena19} for a discussion on the size of the CO emission in ASPECS LP 3\,mm). 

We rebin the frequency dimension in channels of 7.813\,MHz, i.e., $2\times$ the native spectral resolution of the observations. At 99.5\,GHz, this corresponds to $\Delta v\approx 23.5$\,\kms{}. We use `nearest' interpolation scheme in order to maintain the independence of the channels despite the small frequency corrections due to the Earth rotation and revolution during the execution of the observations. We reach a sensitivity of $\sim 0.2$\,mJy\,beam$^{-1}$ per 7.813\,MHz channel throughout the scanned frequency range. For a line width of 200\,\kms{}, these limits correspond to fiducial 5-$\sigma$ CO line luminosity limits of $(1.4, 2.1, 2.3)\times10^9$ \Kkmspc, for CO(2-1), CO(3-2), and CO(4-3), respectively. Via the working assumptions discussed in section \ref{sec_assumptions}, we infer H$_2$ mass limits of $6.8\times10^9$\,\Msun{}, $1.8\times10^{10}$\,\Msun{}, and $2.7\times10^{10}$\,\Msun{} at $1.006<z<1.738$, $2.008<z<3.107$, and $3.011<z<4.475$ respectively. Fig.~\ref{fig_surveys} compares these molecular gas mass limits and volume coverage reached in ASPECS LP 3\,mm with those of all the other molecular scans performed so far. Tab.~\ref{tab_zbin} lists the CO redshift coverage, fiducial gas mass limits, and the volume of universe of ASPECS LP 3\,mm in various CO transitions.

\begin{table}
\caption{\rm CO transitions, redshift bins, cosmic volume, and typical H$_2$ mass limit [at 5-$\sigma$, assuming a line width of 200\,\kms{}, CO excitation as in \citealt{daddi15}, and a CO-to-H$_2$ conversion factor $\alpha_{\rm CO}$=$3.6$ \Msun{} (\Kkmspc)$^{-1}$] in ASPECS LP 3\,mm.} \label{tab_zbin}
\begin{center}
\begin{tabular}{cccc}
\hline
Line    & Redshift      & Volume     & limit $M_{\rm H2}$ \\
        &               & [cMpc$^3$] & [$10^{10}$ M$_\odot$] \\
 (1)    &     (2)       & (3)        & (4)                \\
\hline
CO(1-0) & $0.003-0.369$ &   338  &  0.11 \\   
CO(2-1) & $1.006-1.738$ &  8198  &  0.68 \\   
CO(3-2) & $2.008-3.107$ & 14931  &  1.8  \\   
CO(4-3) & $3.011-4.475$ & 18242  &  2.7  \\   
\hline
\end{tabular}
\end{center}
\end{table}

\section{Ancillary data}\label{sec_ancillary}

The HUDF is one of the best studied extragalactic regions in the sky. Our observations thus benefit from a wealth of ancillary data of unparalleled quality in terms of depth, angular resolution, wavelength coverage, and richness of spectroscopic information. 
When comparing with literature multi-wavelength catalogs, we apply a rigid astrometry offset ($\Delta$RA=$+0.076''$, $\Delta$Dec=$-0.279''$; see \citealt{rujopakarn16,dunlop17}) to available optical/NIR catalogs, in order to account for the different astrometric solution between the ALMA data and optical/NIR data.

The bulk of optical and NIR photometry comes from the {\em Hubble} Space Telescope ({\em HST}) Cosmic Assembly Near-infrared Deep Extragalactic Legacy Survey (CANDELS; \citealt{grogin11}; \citealt{koekemoer11}). These are based both on archival and new {\em  HST} images obtained with the Advanced Camera for Surveys (ACS) at optical wavelengths, and with the Wide Field Camera 3 (WFC3) in the near-infrared. We refer to the photometric compilation by \citet{skelton14}, which also includes ground--based optical and NIR photometry from \citet{nonino09}, \citet{hildebrandt06}, \citet{erben05}, \citet{cardamone10}, \citet{wuyts08}, \citet{retzlaff10}, \citet{hsieh12}, as well as {\em Spitzer} IRAC 3.6\,$\mu$m, 4.5\,$\mu$m, 5.8\,$\mu$m, and 8.0\,$\mu$m photometry from \citet{dickinson03}, \citet{elbaz11}, and \citet{ashby13}. We also include the {\em Spitzer} MIPS 24\,$\mu$m photometric information from \citet{whitaker14}.

The main optical spectroscopy sample in the ASPECS LP footprint comes from the MUSE {\em Hubble} Ultra Deep Survey \citep{bacon17}, a mosaic of nine contiguous fields observed with the Multi Unit Spectroscopic Explorer at the ESO Very Large Telescope. The surveyed area encompasses the entire HUDF. MUSE provides integral field spectroscopy of a $1'\times 1'$ square field over the wavelength range 4750--9300 \AA{}. This yields emission--line redshift coverage in the ranges $z<0.857$, $0.274<z<1.495$, $1.488<z<3.872$, $2.906<z<6.648$ for [O{\sc iii}]$_{5000\,\rm \AA}$, [O{\sc ii}]$_{3727\,\rm \AA}$, C{\sc iii}]$_{1909\,\rm \AA}$, and Ly$\alpha$, respectively. The redshift catalog based on the MUSE {\em Hubble} Ultra Deep Survey consists of $>1500$ galaxies with spectroscopic redshifts in the HUDF \citep{inami17}. We also include any additional spectroscopic information based on various studies at optical and NIR wavelengths, as compiled in \citet{lefevre05}, \citet{coe06}, \citet{skelton14}, and \citet{morris15}.

{\em HST} grism spectroscopy is also available in the HUDF. These observations allow for integral field spectroscopy with sub-arcsec angular resolution at relatively modest ($\lambda/\Delta\lambda\lsim 1000$) spectral resolution. While optical grism spectroscopy of the HUDF has been done \citep{xu07}, we take particular advantage of the more recent {\em HST} grism spectroscopy campaigns at NIR wavelengths, in particular the 3D-HST survey \citep{momcheva16}. This complements the MUSE information, providing spectroscopy of H$\alpha$, H$\beta$ and other rest-frame optical lines at $z=1-3$, together with some additional redshift information in the ``redshift desert'' at $1.5<z<2.9$ where MUSE is less efficient due to the paucity of bright emission lines that are shifted into the MUSE wavelength range at these redshifts.

We create a master catalog of galaxies in the HUDF by combining the \citet{skelton14} catalog with the compilations by \citet{lefevre05}, \citet{coe06}, \citet{xu07},  \citet{rhoads09}, \citet{mclure13}, \citet{schenker13}, \citet{bouwens14,bouwens15}, \citet{morris15}, \citet{inami17}. The catalogs are merged with a simple geometrical association, with an angular threshold of $0.5''$ ($1.0''$) for the photometry (spectroscopy). This selection is also cross-matched with the measurements of morphological parameters (size, ellipticity, light concentration index) from \citet{vanderwel12}. The whole catalog, extending over most of the GOODS--South footprint, consists of $>63000$ entries. In the $2.5'\times2.1'$ area of the XDF, the catalog includes photometry in $>$30 broad and medium bands for $\sim7000$ galaxies, 475 of which have a spectroscopic redshift. 

The photometric dataset is modeled with the high--$z$ extension of the Spectral Energy Distribution (SED) fitting code \textsf{MAGPHYS} \citep{dacunha08,dacunha15}, in order to infer physical parameters: stellar mass, sSFR (and thus, SFR), dust extinction, IR luminosity, etc. We use the available photometry between $0.37$\,$\mu$m and 8.0 $\mu$m, as well as data from the available 1.2\,mm imaging of the field. These results are discussed in detail in \citet{boogaard19}.

\section{Analysis and Results}\label{sec_results}

\begin{figure}
\begin{center}
\includegraphics[width=0.89\columnwidth]{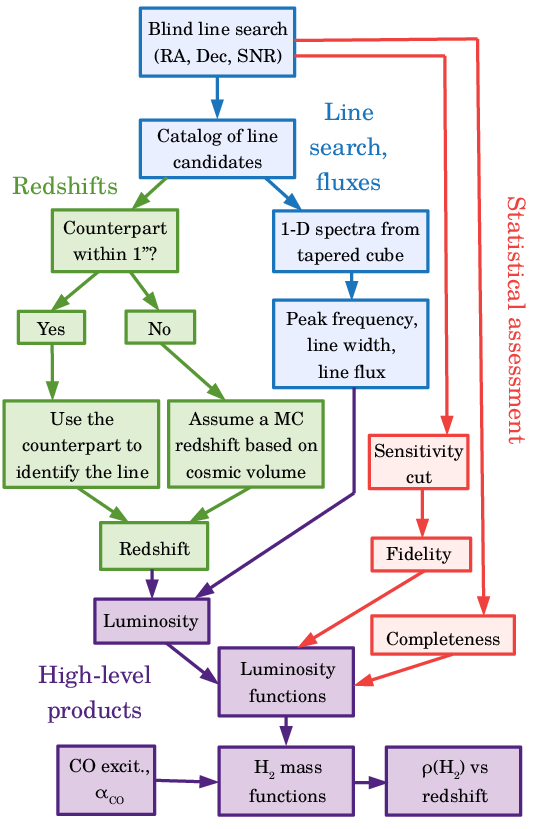}
\end{center}
\caption{A scheme of the workflow followed in this analysis. Four broad areas are identified: The search of line candidates and their characterization in terms of observed quantities (in particular, the line flux), marked in blue; the redshift association, in green; the statistical analysis required to gauge the impact of false positives and of the incompleteness of our search, colored in red; and finally, the high--level data products in purple. }
\label{fig_workflow}
\end{figure}

Our goal is to compute CO luminosity functions and measurements of $\rho$(H$_2$) based on the results from the CO line search in the ASPECS LP 3\,mm data. Our workflow, sketched in Fig.~\ref{fig_workflow}, is articulated in four main blocks: The search for line candidates in the cube, and their characterization in terms of observed quantites (e.g., line fluxes); the assessment of the reliability of the line candidates and of the completeness of our line search; the identification of the line candidates and the measurement of a CO--based redshift; and the construction of high--level data products (e.g., luminosity functions).

\subsection{Line search}

We extensively discuss the line search approach in GL19, and summarize the main steps here for completeness. The cube is searched for emission at any spatial position and spectral coordinate, without any prior based on data from other wavelengths, in order to minimize biases in our selection function. Among the compilations presented in GL19, here we refer to the results obtained with \textsf{findclumps}. This catalog of line candidates consists of 613 entries at S/N$>$5.0, 70 at S/N$>$5.5, 21 at S/N$>$6.0, and 15 at S/N$>$6.5. 

The fidelity or reliability of a line candidate gauges the impact of false positive detections in our search. The idea is to estimate the probability that a given line candidate may be spurious (i.e., a noise feature). The statistics of negative line candidates is used to model the noise properties of the cube, as a function of the S/N and the width of each line candidate\footnote{Since we adopt a matched--filter approach in the line search, the line width affects the reliability of the line candidates in that the narrower the filter kernel, the more independent resolution elements are present in the cube. As a consequence, the probability of finding a high--S/N noise peak increases.}. The fidelity is then defined as 1-$P$, where $P$ is the probability of a (positive) line candidate to be due to noise. We limit our analysis to line candidates with fidelity $>$20\%. We discuss the impact of fidelity on our results in Section~\ref{sec_discussion}. 

The completeness of our line search is estimated by ingesting in the cube mock lines spanning a range of values for various parameters (3D position in the cube, flux, width), under the assumption that the lines are well described by Gaussian profiles. The line search is then repeated, and the completeness is simply inferred as the ratio between the number of retrieved and ingested mock lines, as a function of all the input parameters. In the construction of the CO LFs, we only consider line candidates with a parameter set yielding a completeness $>$20\%.

\begin{figure}
\begin{center}
\includegraphics[width=0.99\columnwidth]{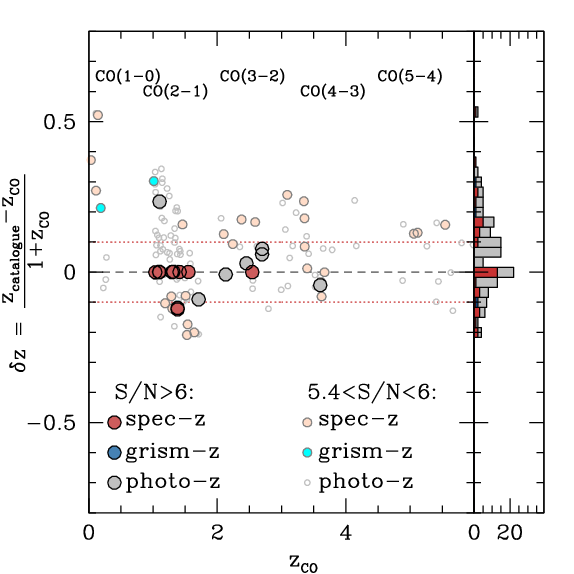}
\end{center}
\caption{Comparison between the CO--based redshifts of the line candidates in our search, and the redshifts available in existing galaxy catalogs in the field. By construction, only line candidates with a tentative counterpart are shown (141 line candidates). The panel on the right shows the collapsed distribution in $\delta z=(z_{\rm cat}-z_{\rm CO})/(1+z_{\rm CO})$. More than half of the sources (79/141) lies within $|\delta z|<0.1$ (dotted dark--red lines). The largest deviations observed in spectroscopically--confirmed redshifts are due to blends of overlapping galaxies along the line of sight (see \citealt{boogaard19}). }
\label{fig_z_match}
\end{figure}

\begin{figure*}
\begin{center}
\includegraphics[height=0.225\textheight]{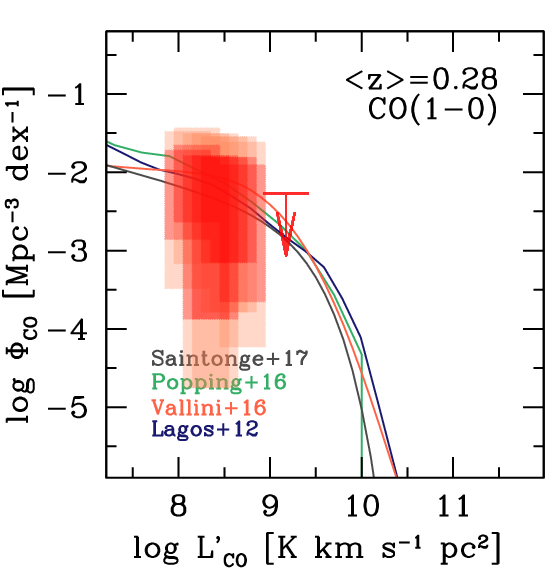}
\includegraphics[height=0.225\textheight]{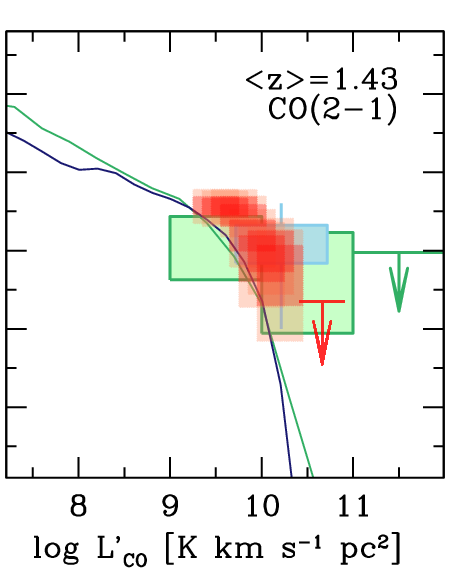}
\includegraphics[height=0.225\textheight]{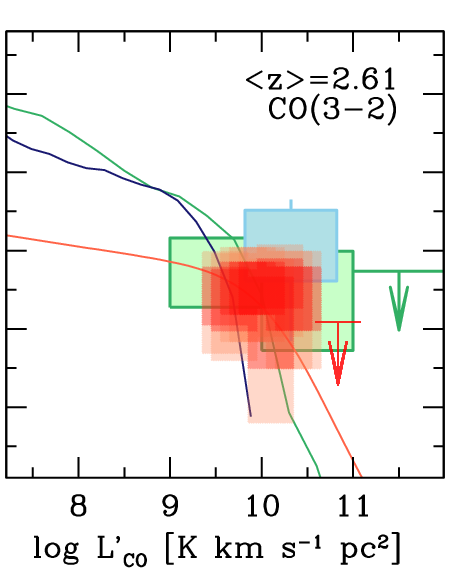}
\includegraphics[height=0.225\textheight]{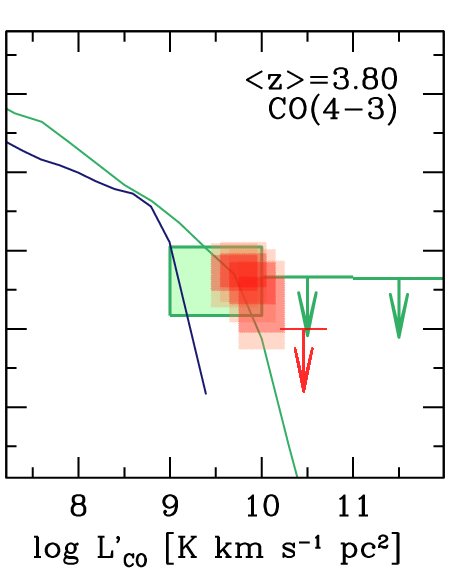}\\
\end{center}
\caption{The ASPECS LP 3\,mm luminosity functions of the observed CO transitions (light red / red shaded boxes, marking the 1-/2-$\sigma$ confidence intervals), compared with the results from the ASPECS Pilot (green boxes; \citealt{decarli16a}), the PdBI HDF-N molecular scan (cyan boxes; \citealt{walter14}), the predicted CO luminosity functions based on the {\em Herschel} IR luminosity functions (red lines; \citealt{vallini16}), and the predictions from semi-analytical models (green lines: \citealt{popping16}; blue lines: \citealt{lagos12}). The ASPECS LP 3\,mm results confirm and expand on the results of the ASPECS Pilot program. We get solid constraints on the CO LFs all the way to $z\approx 4$  (see also \citealt{popping19}). The ASPECS LP 3\,mm results show an excess of bright CO emission compared with the predictions from models.}
\label{fig_co_lf}
\end{figure*}

\begin{figure*}
\begin{center}
\includegraphics[height=0.225\textheight]{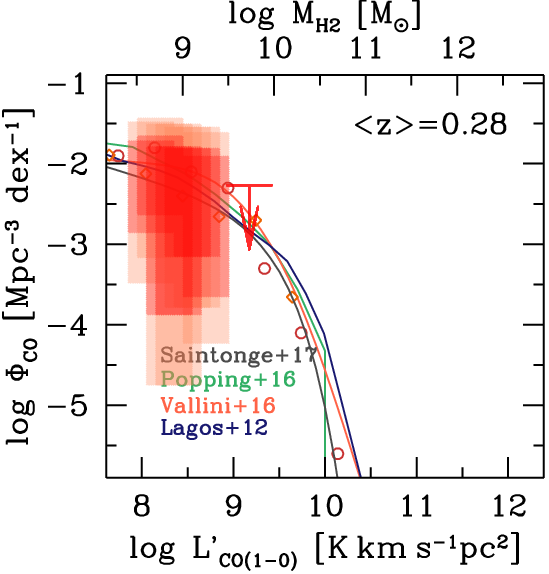}
\includegraphics[height=0.225\textheight]{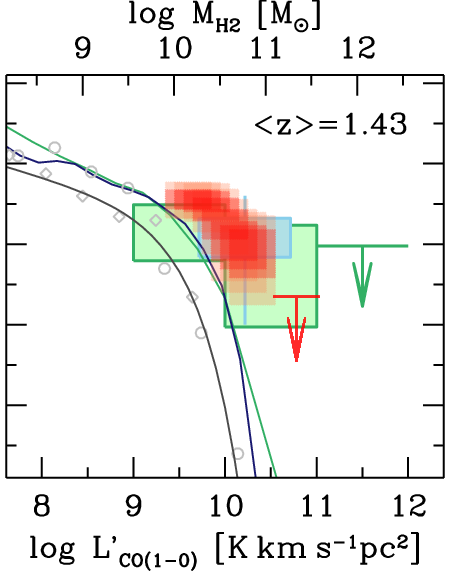}
\includegraphics[height=0.225\textheight]{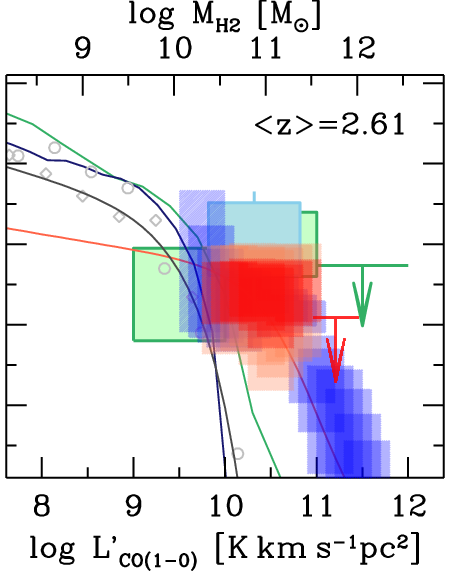}
\includegraphics[height=0.225\textheight]{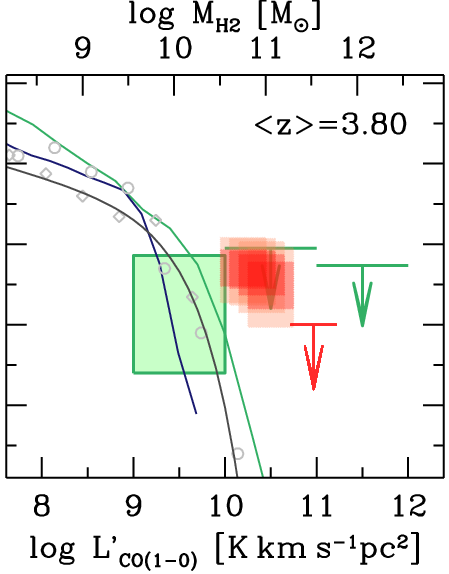}
\end{center}
\caption{Same as Fig.~\ref{fig_co_lf}, but for the corresponding CO(1-0) transition at various redshifts. The CO(1-0) observed LFs at $2<z<3$ from COLDz (blue boxes; \citealt{riechers19}); and the local CO(1-0) LFs (orange diamonds: \citealt{boselli14}; brown circles: \citealt{keres03}; solid grey line: \citealt{saintonge17}. The local constraints are repeated in grey in all the panels for reference). We find strong evidence of an evolution in the CO(1-0) LFs with redshift, with the knee of the CO luminosity function shifting by $>1$ dex towards bright emission between $z\approx 0$ and $z>1$.}
\label{fig_co10_lf}
\end{figure*}

\subsection{Line identification and redshifts}

In order to convert the fluxes of the line candidates into luminosities, we need to identify the observed lines: In principle, the spectral range covered in our 3\,mm scan is broad enough to encompass multiple CO transitions at specific redshifts, thus offering a robust direct constraint on the line identification. However, as shown in Fig.~\ref{fig_nlines_z}, this happens only at relatively high redshifts ($z\gsim3$, if one considers both CO and \Ci{}). We therefore need to consider different approaches to pin down the redshift of our line candidates. First, we search for a counterpart at optical/NIR wavelengths. If successful, we use the available redshift of the counterpart to associate line candidates and CO transitions: if the counterpart has a redshift $z_{\rm cat}<0.8$, $0.8<z_{\rm cat}<1.9$, $1.9<z<3.2$, etc, we identify the line candidate as CO(1-0), CO(2-1), CO(3-2), etc, respectively. The majority of the 21 line candidates with S/N$>$6 show very good agreement ($|\delta z|=|(z_{\rm cat}-z_{\rm CO})/(1+z_{\rm CO})|\lsim 0.01$) between CO--based and catalog redshifts (see Fig.~\ref{fig_z_match}). Other line candidates have a CO redshift roughly consistent ($|\delta z|<0.3$) with the catalog photometric redshifts. Two galaxies detected at S/N$>$6 in CO have a spectroscopic catalog redshift that is inconsistent with the CO--based redshift. Our detailed analysis of the MUSE data confirms that these cases are examples of overlapping galaxies at different redshifts; i.e., both the catalog values and the CO--based values are confirmed \citep{boogaard19}. Fig.~\ref{fig_z_match} shows the comparison between the CO--based and catalog redshifts.

If the line candidates do not have a counterpart at other wavelengths (about 25\% of line candidates at S/N$>$5), the line identification is performed through a bootstrap, where the probability of a line candidate to be CO(1-0), CO(2-1), CO(3-2), and CO(4-3) is proportional to the volume of universe sampled in each of these transitions with ASPECS LP at 3\,mm. We do not consider transitions at higher $J$ values, since significant CO excitation would have to be invoked in order to explain bright high--$J$ line emission. In Appendix \ref{sec_robustness}, we discuss the impact of these assumptions on our results. 

In the construction of CO luminosity functions, we only use CO--based redshifts.

\subsection{Line luminosities and corresponding H$_2$ mass}\label{sec_assumptions}

The line fluxes are transformed into luminosities following \citet{carilli13}:
\begin{equation}\label{eq_linelum}
\frac{L'}{\rm K\,km\,s^{-1}\,pc^2}=\frac{3.257\times10^7}{1+z} \, \frac{F_{\rm line}}{\rm Jy\,km\,s^{-1}} \, \left(\frac{\nu_{\rm 0}}{\rm GHz}\right)^{-2} \left(\frac{D_{\rm L}}{\rm Mpc}\right)^2
\end{equation}
where $F_{\rm line}$ is the integrated line flux, $\nu_0$ is the rest-frame frequency of the line, and $D_{\rm L}$ is the luminosity distance. We then infer the corresponding CO(1-0) luminosities by adopting the CO[$J$-($J$-1)]--to--CO(1-0) luminosity ratios, $r_{J1}$, from \citet{daddi15}: $L'$ [CO(1-0)] = $L' / r_{J1}$, with $r_{J1}=\{1.00$, $0.76\pm0.09$, $0.42\pm0.07$, $0.31\pm0.07\}$,  for $J_{\rm up}$=$\{$1, 2, 3, 4$\}$. These values are based on VLA and PdBI observations of multiple CO transitions in 4 main sequence galaxies at $z\approx1.5$. These galaxies are less extreme than the typical, high IR luminosity galaxies studied in multiple CO transitions at $z>1$, thus likely more representative of the galaxies studied here. We include a bootstrapped realization of the uncertainties on $r_{J1}$ in the conversion. In Appendix~\ref{sec_robustness} we discuss the impact of the $r_{J1}$ assumptions on our results. 

The cosmic microwave background at high redshift enhances the minimum temperature of the cold ISM, and suppresses the observability of CO lines in galaxies because of the lower constrast against the background \citep[for extended discussions, see e.g.,][]{dacunha13,tunnard16}. The net effect is that the observed CO emission is only a fraction of the intrinsic one, with the suppression being larger for lower J transitions and at higher redshifts. This correction is however typically small at $z=1-3$, and often neglected in the literature \citep[e.g.,][]{tacconi18}. Indeed, for $T_{\rm kin}\approx T_{\rm dust}$, and following the $T_{\rm dust}$ evolution in \citet{magnelli14}, we find $T_{kin}>30$\,K at $z>1$, thus yielding CO flux corrections of $\lsim 15$\% up to $z$=$4.5$. Because of its minimal impact, the associated uncertainties, and for consistency with the literature, we do not correct our measurements for the cosmic microwave background impact.

The resulting CO(1-0) luminosities are converted into molecular gas masses, $M_{\rm H2}$, via the assumption of a CO--to--H$_2$ conversion factor, $\alpha_{\rm CO}$:
\begin{equation}\label{eq_Mh2}
M_{\rm H2}=\frac{\alpha_{\rm CO}}{r_{J1}} \, L'
\end{equation}
A widespread assumption in the literature on ``normal'' high--redshift galaxies \citep[e.g.,][]{daddi10a, magnelli12, carilli13, tacconi13, tacconi18, genzel15, riechers19} is a value of $\alpha_{\rm CO}\approx4$ \Msun{}\,(K\,\kms{}\,pc$^2$)$^{-1}$, consistent with the Galactic value \citep[see, e.g.,][]{bolatto13}, once the Helium contribution ($\sim 36$\%) is removed. Here we adopt $\alpha_{\rm CO}=3.6$ \Msun{}\,(K\,\kms{}\,pc$^2$)$^{-1}$ \citep{daddi10a}. 
A different, yet constant choise of $\alpha_{\rm CO}$ would result in a linear scaling of our results involving $M_{\rm H2}$ and $\rho$(H$_2$). This is further discussed in Sec.~\ref{sec_discussion}.

\subsection{CO luminosity functions}

The CO luminosity functions are constructed in a similar way as in \citet{decarli16a} via a Monte Carlo approach that allows us to simultaneously account for all the uncertainties in the line flux estimates, in the line identification, in the conversion factors, as well as for the fidelity of the line candidates. For each line candidate, we compute the corresponding values of completeness and fidelity, based on the observed line properties (S/N, line width, flux, etc). If the line has been confirmed by, e.g., a counterpart with a matching spectroscopic redshift, we assume that the fidelity is 1. In all the other cases, we conservatively treat our fidelity estimates as upper limits, and adopt a random value of fidelity that is uniformly distributed between 0 and such upper limit (see GL19; \citealt{pavesi18}; \citealt{riechers19}). We extract a random number for each entry; line candidates are kept in our analysis only if the random value is below the fidelity threshold (so that, the lower the fidelity, the lower the chances that the line candidate is kept in our analysis). Typically, 20--40 line candidates survive this selection in each realization. 

We split the list of line candidates by CO transitions and in 0.5\,dex wide bins of luminosity. In each bin, we compute the Poissonian uncertainties. We then scale up each entry by the inverse of the completeness. The completeness--corrected entry counts in each bin are then divided by the comoving volume covered in each transition. This is computed by counting the area with sensitivity $>$50\% of the peak sensitivity obtained at the center of the mosaic in each channel. 

The CO luminosity functions are created 1000 times (both for the observed CO transitions, and for the corresponding J=1$\rightarrow$0 ground transition), each time with a different realization of all the parameters that are left uncertain (the fidelity and its error bars, the identification of lines without counterparts, the $r_{J1}$ ratio, etc). The analysis is then repeated five times after a shift of 0.1\,dex of the luminosity bins, which allows us to remove the dependence of the reconstructed CO luminosity functions from the bin definition. The final CO luminosity functions are the averages of all the CO LF realizations. The CO and CO(1-0) LFs are listed in Tables~\ref{tab_co_lf} and \ref{tab_co10_lf} and plotted in Fig.~\ref{fig_co_lf} and \ref{fig_co10_lf}.

The H$_2$ mass functions in our analysis are simply obtained by scaling the CO(1-0) LFs by the (fixed) $\alpha_{\rm CO}$ factor. We then sum the CO--based completeness--corrected H$_2$ masses of each line candidate passing the fidelity threshold in bins of redshift, and we divide by the comoving volume in order to derive the cosmic gas molecular mass density, $\rho$(H$_2$). By construction, we do not extrapolate towards low CO luminosities / low H$_2$ masses. However, in the following we will show that accounting for the faint end would only very marginally affect our results.

\subsection{Analytical fits to the CO LFs}\label{sec_schechter}

We fit the observed CO luminosity functions with a Schechter function \citep{schechter76}, in the logarithmic form used in \citet{riechers19}:
\begin{equation}\label{eq_schechter}
\log \Phi(L')=\log \Phi_* + \alpha \, \log \left(\frac{L'}{L'_*}\right) \, \frac{1}{\ln 10} \frac{L'}{L'_*} + \log(\ln(10))
\end{equation}
where $\Phi(L')\,d(\log L')$ is the number of galaxies per comoving volume with a CO line luminosity between log~$L'$ and log~$L'+d(\log L')$; $\Phi_*$ is the scale number of galaxies per unit volume; $L'_*$ is the scale line luminosity which sets the knee of the luminosity function; $\alpha$ is the slope of the faint end. We fit the observed CO LFs in the three redshift bins at $z>1$ considered in this study; the $z<0.37$ bin is ignored because of the modest luminosity range and sample size in our study. The LFs presented in this work are created in bins of 0.5\,dex spaced by 0.1\,dex, i.e., consecutive bins are not independent. In order to account for this, and to minimize the impact of our bin assumptions, we first fit the LFs using all the available bins, then we repeat the fits on the five independent contiguous subsets of the luminosity bins. 

\begin{figure*}
\begin{center}
\includegraphics[height=0.245\textwidth]{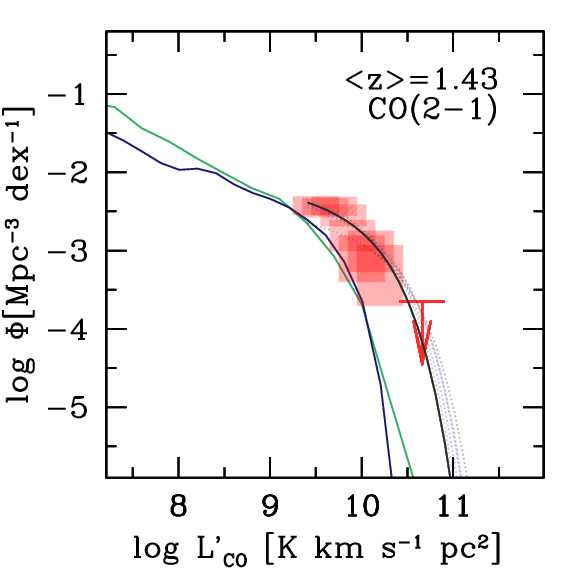}
\includegraphics[height=0.245\textwidth]{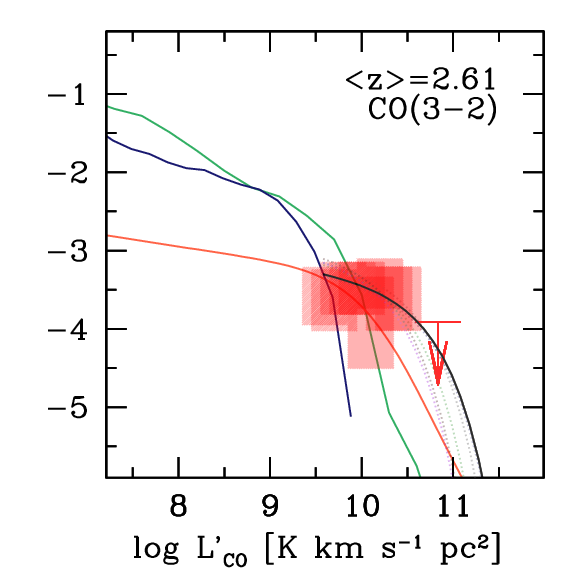}
\includegraphics[height=0.245\textwidth]{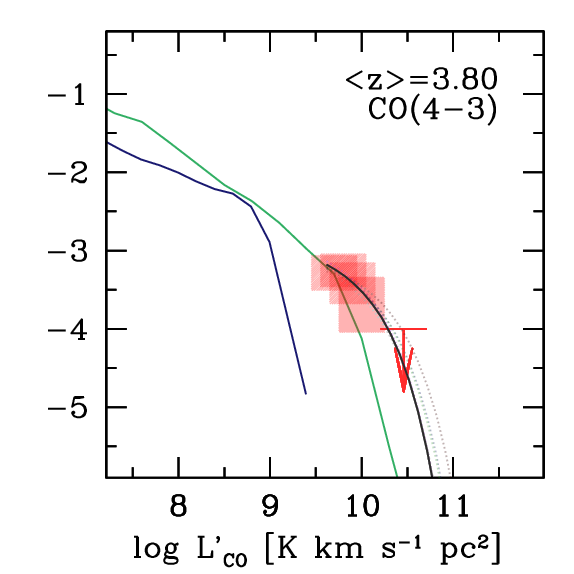}
\end{center}
\caption{Observed CO LFs (red boxes) and their analytical Schechter fits (lines). The best fit obtained by using all the bins is shown with a solid thick line, while the fits obtained via independent subsets of the data are shown in dotted lines. The use of different bins only marginally affects the fits. We find evidence of an increased value of the characteristic luminosity, $L'_*$, at $z\sim2.5$. The panels also show predictions from the semi-analytical models by \citet{lagos11} and \citet{popping14} (blue and green solid lines, respectively). 
}
\label{fig_co_lf_fits}
\end{figure*}

The slope of the faint end of the LF, $\alpha$, is very sensitive to the corrections we apply for fidelity and completeness (see previous section). We therefore opt to conservatively use a fiducial fixed value of $\alpha$=--0.2 in our analysis. This is consistent with findings at $z\approx 0$ \citep[][once we take into account the different definition of $\alpha$]{saintonge17}, as well as with the typical slope of the stellar mass function of field galaxies at various redshifts \citep[e.g.,][]{ilbert13}. As for the other two parameters, we assume broad ($\sigma$=$0.5$\,dex) log normal distributions as priors in $\Phi_*$ and $L'_*$, centered around $10^{-3}$\,Mpc$^{-3}$\,dex$^{-1}$, and $10^{9.5}$\,K\,\kms\,pc$^2$, respectively. The best fit value and the 1-$\sigma$ confidence levels of the fitted parameters are derived from the 50\%, 14\%, and 86\% quartiles of the marginalized posterior distributions of each parameter. They are listed in Table \ref{tab_lf_fits}. Fig.~\ref{fig_co_lf_fits} compares the observed CO LFs with the fitted Schechter functions. 

The fitted parameters do not show strong dependency on the choice of binning, with the results being typically consistent within 1-$\sigma$ uncertainties. We find an indication of a higher $L'_*$ at $z$=1--3 (log $L'_*$~[\Kkmspc] $\approx 10.4$) compared to the $z>3$ bin, and most importantly, with the local universe (log $L'_*$~[\Kkmspc] $\approx 9.9$, although with a different definition of the Schechter function; \citealt{saintonge17}). 

The luminosity--weighted integral of the fitted LFs suggests that the ASPECS LP 3\,mm data recover 83\%, 91\%, and 71\% of the total CO luminosity at $\langle z\rangle$ = 1.43, 2.61, and at 3.80. In addition, if we adopt the best fit by \citet{saintonge17} for the lowest redshift bin, ASPECS LP 3\,mm recovers 59\% of the total CO(1-0) luminosity in the local universe, although this last measure is strongly affected by cosmic variance due to the small volume probed by ASPECS LP 3\,mm.

\begin{table}
\caption{\rm Results of the Schechter fits of the observed CO LFs, assuming a fixed $\alpha=-0.2$.} \label{tab_lf_fits}
\begin{center}
\begin{tabular}{ccc}
\hline
Line    & log $\Phi_*$              & log $L'_*$	\\
        & [Mpc$^{-3}$\,dex$^{-1}$]  & [K\,\kms\,pc$^2$] \\
 (1)    &     (2)                   & (3)		\\
\hline
\multicolumn{3}{c}{\em All $L'$ bins} \\
CO(2-1) & $-2.79_{-0.09}^{+0.09}$ & $10.09_{-0.09}^{+0.10}$  \\
CO(3-2) & $-3.83_{-0.12}^{+0.13}$ & $10.60_{-0.15}^{+0.20}$  \\
CO(4-3) & $-3.43_{-0.22}^{+0.19}$ & $ 9.98_{-0.14}^{+0.22}$  \\
\hline
\multicolumn{3}{c}{\em Independent $L'$ bins } \\
CO(2-1) & $ -2.93_{-0.12}^{+0.11} $ & $ 10.23_{-0.11}^{+0.16} $ \\
CO(2-1) & $ -2.90_{-0.14}^{+0.16} $ & $ 10.22_{-0.22}^{+0.24} $ \\
CO(2-1) & $ -2.77_{-0.20}^{+0.21} $ & $ 10.12_{-0.25}^{+0.35} $ \\
CO(2-1) & $ -2.86_{-0.14}^{+0.15} $ & $ 10.17_{-0.17}^{+0.17} $ \\
CO(2-1) & $ -3.14_{-0.19}^{+0.19} $ & $ 10.32_{-0.18}^{+0.26} $ \\
        &  	                    &  	     	     	        \\
CO(3-2) & $ -3.65_{-0.23}^{+0.25} $ & $ 10.49_{-0.22}^{+0.26} $ \\
CO(3-2) & $ -3.85_{-0.20}^{+0.21} $ & $ 10.59_{-0.20}^{+0.23} $ \\
CO(3-2) & $ -3.63_{-0.17}^{+0.17} $ & $ 10.36_{-0.21}^{+0.25} $ \\
CO(3-2) & $ -3.55_{-0.26}^{+0.28} $ & $ 10.22_{-0.21}^{+0.19} $ \\
CO(3-2) & $ -3.50_{-0.21}^{+0.22} $ & $ 10.24_{-0.15}^{+0.21} $ \\
        &  	                    &  	     	     	        \\
CO(4-3) & $ -3.53_{-0.28}^{+0.36} $ & $ 10.01_{-0.21}^{+0.26} $ \\
CO(4-3) & $ -3.55_{-0.26}^{+0.23} $ & $ 10.10_{-0.16}^{+0.18} $ \\
CO(4-3) & $ -3.53_{-0.19}^{+0.18} $ & $ 10.08_{-0.15}^{+0.20} $ \\
CO(4-3) & $ -3.38_{-0.28}^{+0.26} $ & $  9.98_{-0.20}^{+0.36} $ \\
CO(4-3) & $ -3.59_{-0.23}^{+0.25} $ & $ 10.21_{-0.25}^{+0.40} $ \\
\hline
\end{tabular}
\end{center}
\end{table}

\section{Discussion}\label{sec_discussion}

\begin{figure*}
\begin{center}
\includegraphics[width=0.79\textwidth]{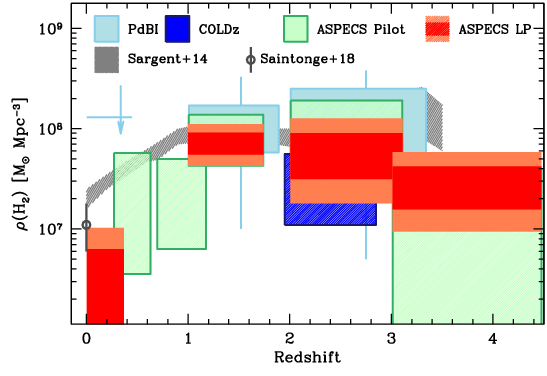}
\end{center}
\caption{The redshift evolution of the cosmic molecular gas density, $\rho$(H$_2$), as constrained by ASPECS LP 3\,mm (red shaded regions) and by other molecular scans: the PdBI scan \citep{walter14}, the COLDz survey \citep{riechers19}, and the ASPECS Pilot \citep{decarli16a} (shown in cyan, blue, and green boxes respectively), compared with the local measure by \citet{saintonge17} (grey circle). The grading in the ASPECS LP boxes highlight the 1-, and 2-$\sigma$ confidence levels. The ASPECS LP 3\,mm constraint on $\rho$(H$_2$) at $z<0.3$ is below the estimates from local studies, likely due the small $L'_{\rm CO}$ range sampled in ASPECS LP 3\,mm, and the higher impact of cosmic variance due to the small volume we probed. Our new data show that the molecular gas content slowly increases from early cosmic epochs up to $z\sim 1.5$, then dropped by a factor $\sim 6$ to the present day. This is fully consistent with the constraints derived from other molecular scans, irrespective of the region of the sky they surveyed. The evolution appears more pronounced than what most semi-analytical models predict (see, e.g., \citealt{lagos11}, \citealt{popping14}, and the discussion in \citealt{popping19}). The observed evolution in $\rho$(H$_2$) seems to closely match the evolution in $\rho_{\rm SFR}$ \citep{madau14}, thus suggesting that the gas content is the main driver of the star formation history.}
\label{fig_rhoH2}
\end{figure*}

Figs.~\ref{fig_co_lf} and \ref{fig_co10_lf} show that ASPECS LP 3\,mm sampled a factor $\sim 20$ in CO luminosity at $z=1-4$ (see also Tables \ref{tab_co_lf} and \ref{tab_co10_lf}). We find evidence of an evolution in the CO LFs [and in the corresponding CO(1-0) LFs] as a function of redshift, compared to the local universe \citep{keres03,boselli14,saintonge17}, suggesting that the characteristic CO luminosity of galaxies at $z$=1--4 is an order of magnitude higher than in the local universe, once we account for CO excitation. This is in line with the findings from other studies, e.g., other molecular scans \citep{walter14,decarli16a,riechers19}; targeted CO observations on large samples of galaxies \citep[e.g.,][]{genzel15,aravena16c,tacconi18}; and similar works based on dust continuum observations \citep[e.g.,][]{magnelli13,gruppioni13,scoville17}. The CO LFs show an excess at the bright end compared with the predictions by semi--analytical models \citep{lagos11,popping14}, and more compatible with empirical predictions \citep{sargent14,vallini16}. Fig.~\ref{fig_rhoH2} demonstrates that a prominent evolution in $\rho$(H$_2$) occurred between $z\approx4$ and nowadays, with the molecular gas content in galaxies slowly rising since early cosmic epochs, peaking around $z$=1--3, and dropping by a factor $6.5_{-1.4}^{+1.8}$ down to the present age (see also Table \ref{tab_rhoH2}). The values of $\rho$(H$_2$) used here only refer to the actual line candidates, i.e., we do not attempt to extrapolate towards undetected faint end of the LFs. However, as discussed in Sec.~\ref{sec_schechter}, our observations recover close to 90\% of the total CO luminosity at $z=1.0$--$3.1$ (under the assumption of a slope of $\alpha$=--$0.2$ for the faint end), i.e., the derived $\rho$(H$_2$) values would shift upwards by small factors ($\sim 10-20$\%). In Appendix \ref{sec_robustness}, we test the robustness of the CO LFs and $\rho$(H$_2$) evolution with redshift against some of the working assumptions in our analysis. A different choice of $\alpha_{\rm CO}$ would linearly affect our results on $\rho$(H$_2$). In particular, by adopting $\alpha_{\rm CO}\approx 2$ \Msun{}\,(K\,\kms{}\,pc$^2$)$^{-1}$, as the comparison between dust--based and CO--based gas masses suggests \citep{aravena19}, we would infer a milder evolution of $\rho$(H$_2$) at $z>1$ and the local measurements. 

In the following, we discuss our results in the context of previous studies.

\subsection{CO Luminosity functions}

Compared to any previous molecular scan at mm wavelengths \citep{walter14, decarli16a, riechers19}, ASPECS LP 3\,mm provides superior sample statistics, which enables the more detailed analysis described in this series of papers. As shown in Fig.~\ref{fig_surveys}, ASPECS LP 3\,mm complements very well COLDz in that it samples a smaller volume but reaching a deeper sensitivity. The large volumes sampled by ASPECS LP 3\,mm and COLDz, and the different targeted fields, mitigate the impact of cosmic variance. Overall, the CO LFs observed from the ASPECS LP 3\,mm data appear in good agreement with the constraints from the first molecular scan observations (see Fig.~\ref{fig_co_lf}). 

Fig.~\ref{fig_co_lf} compares our observed CO LFs with the LF predictions by the semi-analytical models presented in \citet{lagos11} and \citet{popping14}. Semi-analytical models tend to underpredict the bright end of the CO luminosity functions at $z>1$, with a larger discrepancy for the \citet{lagos11} models at $z>2$. The tension increases if we compare our inferred CO(1-0) LFs with the predictions from models (Fig.~\ref{fig_co10_lf}). This hints at an intrinsic difference on how widespread large molecular gas reservoirs in high--redshift galaxies are as predicted by models, with respect to that suggested by our observations; the tension is somewhat reduced by the different treatment of the CO excitation (see Appendix~\ref{sec_robustness}).

The CO(1-0) LF inferred in our study at $2.0<z<3.1$ is in excellent agreement with the one derived from the COLDz survey \citep{riechers19} (see Fig.~\ref{fig_co10_lf}). Because of the different parameter space, the COLDz data show larger uncertainties in the faint end, but provide a better constraint on the bright end compared to ASPECS LP 3\,mm. The good match between these two independent observations might be considered as supporting evidence that the impact of cosmic variance is relatively modest (the targeted fields are completely independent); and that our assumption on the CO excitation, used to transform CO(3-2) into its corresponding CO(1-0), works reasonably well. Interestingly, the CO(1-0) LFs from ASPECS LP 3\,mm and COLDz derived at $z\sim2.5$ are in good agreement with the empirical predictions by \citet{vallini16}, based on the {\em Herschel} IR luminosity functions \citep{gruppioni13}. 

It is also interesting to compare the inferred CO(1-0) LFs with the ones measured in the local universe by \citet{keres03}, \citet{boselli14}, and \citet{saintonge17}. The local measurements differ from each other by up to a factor $\sim 2$. Nevertheless, the ASPECS LP 3\,mm data show a very clear evolution in the CO(1-0) LFs, with a shift upward of the knee of the luminosity functions by an order of magnitude, or an excess by several orders of magnitudes in the number density of bright ($L'>10^{10}$\,K\,\kms\,pc$^2$) CO(1-0) emitters at $z>1$ compared to the local universe.

\subsection{$\rho$(H$_2$) vs redshift}

Fig.~\ref{fig_rhoH2} compares the observed evolution of $\rho$(H$_2$) as a function of redshift from the available molecular scan efforts. The ASPECS LP 3\,mm data confirm the results from the PdBI scan in the HDF-N \citep{walter14} and from the ASPECS Pilot \citep{decarli16a}, but with much tighter constraints thanks to the superior statistics. The cosmic density of molecular gas in galaxies appears to increase by a factor $6.5_{-1.4}^{+1.8}$ from the local universe [$\rho$(H$_2$)$\approx$1.1$\times$10$^7$\,\Msun{}\,Mpc$^{-3}$; \citealt{keres03}, \citealt{boselli14}, \citealt{saintonge17}] to $z\sim 1$ [$\rho$(H$_2$)$\approx$7.1$\times$10$^7$\,Mpc$^{-3}$]\footnote{The ASPECS LP 3\,mm constraint on $\rho$(H$_2$) at $z<0.3$ is below the estimates from local studies; this is likely due to the fact that we only sampled a small luminosity range in $L'_{\rm CO}$ in this redshift bin, in a tiny cosmic volume; furthermore, the HUDF was originally chosen to be relatively underdense of nearby galaxies.}, then follows a relatively flat evolution or possibly a mild decline towards higher redshifts. This is in excellent agreement with the constraints on $\rho$(H$_2$) from COLDz \citep{riechers19} at $2.0<z<2.8$, and with the empirical predictions derived by \citet{sargent14} based on the ``2--star formation mode'' framework, where the distributions of various galaxy properities (gas fraction, star formation efficiency, metallicity, etc) are inferred based on empirical relations, with a key role due to the offset of galaxies with respect to the ``main sequence''. This analysis results in a similar evolution of $\rho$(H$_2$) with redshift as the one found in ASPECS LP 3\,mm. 

The observed evolution in $\rho$(H$_2$) is also in qualitative agreement with other observational studies. E.g., most studies searching for CO emission in targeted observations of main sequence galaxies find that $z$=1--3 galaxies typically have 5--10 times more gas than galaxies of similar stellar mass in the local universe \citep[see, e.g.,][]{genzel15,schinnerer16,tacconi18}. This is in line with the ASPECS LP 3\,mm results, which point to a larger molecular gas content in typical galaxies at $z>1$ \citep[see also][]{aravena19}. 
A similar trend is also reported by studies tackling the problem using dust as a probe of the gas content in high--$z$ galaxies \citep[e.g.][]{magnelli13,gruppioni13}. E.g., \citet{scoville17} put indirect constraints on $\rho$(H$_2$) at various redshifts using dust continuum measurements of {\em Herschel}--selected galaxies, and scaling by an internally--calibrated dust--to--gas ratio. The evolution of $\rho$(H$_2$) that \citet{scoville17} infer is qualitatively similar, although somewhat shallower than the one observed in ASPECS LP 3\,mm, spanning only a factor $\sim 2.5$ in $\rho$(H$_2$) compared to a factor $\sim 6.0$ found in ASPECS LP 3\,mm.

\section{Conclusions}\label{sec_conclusions}

We presented the ASPECS LP 3\,mm survey, an ALMA molecular scan encompassing most of the {\em Hubble} XDF over a large fraction of the 3\,mm transparent band of the atmosphere. We exploited our data to search for massive molecular gas reservoirs (as traced by CO emission) in galaxies throughout $\sim 90$\% of cosmic history, with no prior on counterparts at other wavelengths. We detected 70 line candidates with S/N$>$5.5, $>$75\% of which with a photometric counterpart at optical/NIR wavelengths. This search allowed us to put stringent constraints on the CO luminosity functions in various redshift bins, as well as to infer the cosmic density of molecular gas in galaxies, $\rho$(H$_2$). We found that:

\begin{itemize} 
\item[{\em i-}] The CO luminosity functions undergo significant evolution compared to the local universe. High redshift galaxies appear brighter in CO than galaxies in the local universe. In particular, at $z$=1--3, the characteristic CO(2-1) and CO(3-2) luminosity is $>3\times$ higher than the characteristic CO(1-0) luminosity observed in the local universe. The evolution is even stronger if we account for CO excitation. Analytical fits of our results suggests that we recovered the majority (up to 90\%, depending on assumptions on the faint end) of the total CO luminosity at $z$=1.0--3.1.
\item[{\em ii-}] Similarly, $\rho$(H$_2$) shows a clear evolution with cosmic time: It slowly increased since early cosmic epochs, reached a peak around $z$=1--3, and then decreased by a factor $6.5_{-1.4}^{+1.8}$ to the present day. This factor changes if $\alpha_{\rm CO}$ is allowed to evolve with redshift. In particular, the factor would be $\sim$3 if we adopt $\alpha_{\rm CO}$=2 \Msun{} (\Kkmspc)$^{-1}$ for galaxies at $z>1$.
\item[{\em iii-}] Our results are in agreement with those of other molecular scans which targeted different regions of the sky and sampled different parts of the parameter space (in terms of depth, volume, transitions, etc). Similarly, we generally confirm empirical predictions based on dust continuum observations and SED modeling.
\item[{\em iv-}] Our results are in tension with predictions by semi-analytical models, which struggle to reproduce the bright end of the observed CO LFs. The discrepancy might be mitigated with different assumptions on the CO excitation and $\alpha_{\rm CO}$. \citet{popping19} quantitatively address the comparison between models and the ASPECS LP 3\,mm observations and the underlying assumptions of both.
\item[{\em v-}] Our results hold valid if we restrict our analysis to the subset of galaxies with counterparts at redshifts that strictly match those inferred from our CO observations. The results are qualitatively robust against different assumptions concerning the CO excitation. 
\end{itemize}

The observed evolution of $\rho$(H$_2$) is in quantitative agreement with the evolution of the cosmic star formation rate density \citep[$\rho_{\rm SFR}$; see, e.g.,][]{madau14}, which also shows a mild increase up to $z$=1--3, followed by a drop by a factor $\approx$8 down to present day. Given that the star formation rate can be expressed as the product of the star formation efficiency (= star formation per unit gas mass) and the gas content mass, the similar evolution of $\rho$(H$_2$) and $\rho_{\rm SFR}$ leaves little room for a significant evolution of the star formation efficiency throughout 85\% of cosmic history ($z\approx3$), at least when averaged over the entire galaxy population. The history of cosmic star formation appears dominated by the evolution in the molecular gas content of galaxies.

\acknowledgements

We thank the anonymous referee for their useful feedback which allowed us to improve the quality of the paper.
Este trabajo cont\'o con el apoyo de CONICYT + Programa de Astronom\'ia+ Fondo CHINA-CONICYT. J.G.L.\ acknowledges partial support from ALMA-CONICYT project 31160033. D.R.\ acknowledges support from the National Science Foundation under grant number AST-1614213. F.E.B.\ acknowledges support from CONICYT-Chile Basal AFB-170002 and the Ministry of Economy, Development, and Tourism's Millennium Science Initiative through grant IC120009, awarded to The Millennium Institute of Astrophysics, MAS. I.R.S.\ acknowledges support from the ERC Advanced Grant DUSTYGAL (321334) and STFC (ST/P000541/1). T.D-S.\ acknowledges support from ALMA-CONICYT project 31130005 and FONDECYT project 1151239. J.H.\ acknowledges support of the VIDI research programme with project number 639.042.611, which is (partly) financed by the Netherlands Organisation for Scientific Research (NWO). 

\facility{ALMA} data: 2016.1.00324.L. ALMA is a partnership of ESO (representing its member states), NSF (USA) and NINS (Japan), together with NRC (Canada), NSC and ASIAA (Taiwan), and KASI (Republic of Korea), in cooperation with the Republic of Chile. The Joint ALMA Observatory is operated by ESO, AUI/NRAO and NAOJ.

\appendix

\section{Measured CO Luminosity Functions}

For the sake of reproducibility of our results, Table \ref{tab_co_lf} reports the measured CO LFs in ASPECS LP 3\,mm. Similarly, Table \ref{tab_co10_lf} provides the inferred CO(1-0) LFs from this study. Table \ref{tab_rhoH2} lists the estimated values of $\rho$(H$_2$) in various redshift bins and under different working hypothesis (see Appendix~\ref{sec_robustness}). Finally, Table \ref{tab_catalog} lists the entries of the line candidates used in the construction of the LFs. 
%The full table is provided in the on-line version of the paper.

\begin{table*}
\caption{\rm Luminosity functions of the observed CO transitions. (1, 5) Luminosity bin center; each bin is 0.5\,dex wide. (2-4, 6-8) CO luminosity functions, reported as the minimum and maximum values of the confidence levels at 1, 2, and 3-$\sigma$.} \label{tab_co_lf}
\begin{center}
\begin{tabular}{ccc|ccc}
\hline
log $L'$ & log $\Phi$, 1-$\sigma$ & log $\Phi$, 2-$\sigma$ & log $L'$ & log $\Phi$, 1-$\sigma$ & log $\Phi$, 2-$\sigma$ \\
{}[K\,km\,s$^{-1}$\,pc$^2$] & [dex$^{-1}$\,cMpc$^{-3}$] & [dex$^{-1}$\,cMpc$^{-3}$] & {}[K\,km\,s$^{-1}$\,pc$^2$]  & [dex$^{-1}$\,cMpc$^{-3}$] & [dex$^{-1}$\,cMpc$^{-3}$] \\
 (1)  & (2)          & (3)           & (4)   & (5)          &   (6)           \\
\hline
      & \multicolumn{2}{c}{CO(1-0)}  &       & \multicolumn{2}{c}{CO(2-1)}    \\
 8.0  & -2.86 -1.72  &  -3.48  -1.48 &  9.4  & -2.63 -2.37  &  -2.75 -2.27    \\
 8.1  & -2.86 -1.72  &  -3.48  -1.48 &  9.5  & -2.54 -2.31  &  -2.66 -2.22    \\
 8.2  & -2.64 -1.65  &  -3.17  -1.43 &  9.6  & -2.58 -2.33  &  -2.69 -2.24    \\
 8.3  & -3.88 -1.79  &  -4.75  -1.51 &  9.7  & -2.55 -2.31  &  -2.67 -2.22    \\
 8.4  & -3.88 -1.79  &  -4.75  -1.51 &  9.8  & -2.69 -2.41  &  -2.83 -2.30    \\
 8.5  & -3.16 -1.81  &  -3.79  -1.55 &  9.9  & -3.01 -2.61  &  -3.21 -2.47    \\
 8.6  & -3.15 -1.81  &  -3.78  -1.55 & 10.0  & -3.40 -2.81  &  -3.72 -2.64    \\
 8.7  & -3.65 -1.91  &  -4.23  -1.61 & 10.1  & -3.27 -2.75  &  -3.56 -2.59    \\
 8.8  & -3.65 -1.91  &  -4.23  -1.61 & 10.2  & -3.70 -2.93  &  -4.16 -2.73    \\
 8.9  & -3.65 -1.91  &  -4.23  -1.61 & 10.3  & -3.76 -2.95  &  -4.25 -2.75    \\
 9.0  & -5.52 -1.97  &  -6.39  -1.65 & 10.4  & -3.76 -2.95  &  -4.25 -2.75    \\
      &              &               & 10.5  & -3.76 -2.95  &  -4.25 -2.75    \\
\hline
      & \multicolumn{2}{c}{CO(3-2)}  &       & \multicolumn{2}{c}{CO(4-3)}    \\
 9.6  & -3.95 -3.21  &  -4.31 -3.01  &  9.6  & -3.56 -3.09  &  -3.78 -2.93    \\
 9.7  & -4.03 -3.24  &  -4.41 -3.03  &  9.7  & -3.51 -3.06  &  -3.72 -2.91    \\
 9.8  & -3.82 -3.14  &  -4.16 -2.96  &  9.8  & -3.46 -3.04  &  -3.65 -2.89    \\
 9.9  & -3.82 -3.14  &  -4.16 -2.96  &  9.9  & -3.68 -3.15  &  -3.91 -2.99    \\
10.0  & -3.82 -3.14  &  -4.16 -2.96  & 10.0  & -4.04 -3.34  &  -4.26 -3.13    \\
10.1  & -4.51 -3.34  &  -5.20 -3.11  & 10.1  & -4.04 -3.34  &  -4.26 -3.13    \\
10.2  & -3.74 -3.10  &  -4.10 -2.92  & 10.2  & -4.17 -3.40  &  -4.35 -3.18    \\
10.3  & -4.02 -3.21  &  -4.51 -3.01  & 10.3  & -5.19 -3.59  &  -6.06 -3.31    \\
10.4  & -4.02 -3.21  &  -4.51 -3.01  &       &              &                 \\
\hline
\end{tabular}
\end{center}
\end{table*}

\begin{table*}
\caption{\rm Inferred CO(1-0) luminosity functions in various redshift bins. (1, 4) Luminosity bin center; each bin is 0.5\,dex wide. (2-3, 5-6) CO luminosity functions, reported as the minimum and maximum values of the confidence levels at 1- and 2-$\sigma$.} \label{tab_co10_lf}
\begin{center}
\begin{tabular}{ccc|ccc}
\hline
log $L'$ & log $\Phi$, 1-$\sigma$ & log $\Phi$, 2-$\sigma$ & log $L'$ & log $\Phi$, 1-$\sigma$ & log $\Phi$, 2-$\sigma$ \\
{}[K\,km\,s$^{-1}$\,pc$^2$] & [dex$^{-1}$\,cMpc$^{-3}$] & [dex$^{-1}$\,cMpc$^{-3}$] & {}[K\,km\,s$^{-1}$\,pc$^2$]  & [dex$^{-1}$\,cMpc$^{-3}$] & [dex$^{-1}$\,cMpc$^{-3}$] \\
 (1)  & (2)          & (3)           & (4)   & (5)          &   (6)           \\
\hline
\multicolumn{3}{c}{$0.003<z<0.369$}     & \multicolumn{3}{c}{$1.006<z<1.738$}    \\
 8.0  & -2.86 -1.72  &  -3.48 -1.48  	&  9.5  & -2.66 -2.40  &  -2.79 -2.30	 \\
 8.1  & -2.86 -1.72  &  -3.48 -1.48  	&  9.6  & -2.56 -2.32  &  -2.68 -2.23	 \\
 8.2  & -2.64 -1.65  &  -3.17 -1.43  	&  9.7  & -2.57 -2.33  &  -2.69 -2.23	 \\
 8.3  & -3.88 -1.79  &  -4.75 -1.51  	&  9.8  & -2.60 -2.35  &  -2.72 -2.25	 \\
 8.4  & -3.88 -1.79  &  -4.75 -1.51  	&  9.9  & -2.71 -2.42  &  -2.85 -2.32	 \\
 8.5  & -3.16 -1.81  &  -3.79 -1.55  	& 10.0  & -2.94 -2.57  &  -3.12 -2.44	 \\
 8.6  & -3.15 -1.81  &  -3.78 -1.55  	& 10.1  & -3.20 -2.72  &  -3.46 -2.56	 \\
 8.7  & -3.65 -1.91  &  -4.23 -1.61  	& 10.2  & -3.31 -2.77  &  -3.60 -2.61	 \\
 8.8  & -3.65 -1.91  &  -4.23 -1.61  	& 10.3  & -3.42 -2.82  &  -3.75 -2.65	 \\
 8.9  & -3.65 -1.91  &  -4.23 -1.61  	& 10.4  & -3.55 -2.88  &  -3.94 -2.69	 \\
 9.0  & -5.52 -1.97  &  -6.39 -1.65  	& 10.5  & -3.74 -2.94  &  -4.22 -2.74	 \\
      &              &               	& 10.6  & -3.91 -3.01  &  -4.43 -2.79	 \\
      &              &               	& 10.7  & -4.44 -3.20  &  -5.00 -2.93	 \\
      &              &               	& 10.8  & -5.71 -3.34  &  -6.54 -3.03	 \\
      &              &               	& 10.9  & -6.78 -3.35  &  -7.65 -3.04	 \\
\hline
\multicolumn{3}{c}{$2.008<z<3.107$}     & \multicolumn{3}{c}{$3.011<z<4.475$}    \\
10.0  & -3.99 -3.23  &  -4.36 -3.02  	& 10.1  & -3.58 -3.10  &  -3.79 -2.94    \\
10.1  & -3.97 -3.22  &  -4.33 -3.02  	& 10.2  & -3.52 -3.07  &  -3.73 -2.92    \\
10.2  & -3.96 -3.21  &  -4.32 -3.01  	& 10.3  & -3.55 -3.08  &  -3.76 -2.93    \\
10.3  & -3.90 -3.18  &  -4.26 -2.99  	& 10.4  & -3.63 -3.13  &  -3.85 -2.97    \\
10.4  & -4.07 -3.24  &  -4.50 -3.03  	& 10.5  & -3.80 -3.22  &  -4.03 -3.04    \\
10.5  & -4.29 -3.30  &  -4.81 -3.08  	& 10.6  & -4.01 -3.33  &  -4.23 -3.12    \\
10.6  & -4.15 -3.26  &  -4.65 -3.05  	& 10.7  & -4.27 -3.43  &  -4.49 -3.20    \\
10.7  & -4.01 -3.21  &  -4.46 -3.01  	& 10.8  & -4.63 -3.54  &  -4.85 -3.28    \\
10.8  & -3.96 -3.19  &  -4.40 -3.00  	& 10.9  & -5.24 -3.64  &  -5.50 -3.35    \\
10.9  & -4.01 -3.20  &  -4.48 -3.01  	& 11.0  & -5.86 -3.69  &  -6.11 -3.38    \\
\hline
\end{tabular}
\end{center}
\end{table*}

\begin{table}
\caption{\rm Constraints on $\rho$(H$_2$) in various redshift bins. The quoted ranges correspond to the 1-$\sigma$ and 2-$\sigma$ confidence levels in our analysis. We provide our reference estimates based on the whole sample, and assuming intermediate CO excitation \citep{daddi15} (see Fig.~\ref{fig_rhoH2}), as well as the estimates for the secure sources only (Fig.~B\ref{fig_co_lf_secure_z}) and for the whole sample, but using different assumptions for the CO excitation (Fig.~B\ref{fig_co10_lf_excit}).} \label{tab_rhoH2}
\begin{center}
\begin{tabular}{cccc}
\hline
Redshift       & log $\rho$(H$_2$), 1-$\sigma$ & log $\rho$(H$_2$), 2-$\sigma$  \\
bin            & [\Msun{}\,Mpc$^{-3}$] & [\Msun{}\,Mpc$^{-3}$]  \\
 (1)           &     (2)                  & (3)                      \\
\hline
\multicolumn{3}{c}{\em Reference estimate} \\
0.003--0.369 & 5.89--6.80 & 5.40--7.01  \\ 
1.006--1.738 & 7.74--7.96 & 7.63--8.05  \\ 
2.008--3.107 & 7.50--7.96 & 7.26--8.10  \\ 
3.011--4.475 & 7.20--7.62 & 6.97--7.77  \\ 
\hline
\multicolumn{3}{c}{\em Secure sources only} \\
0.003--0.369 & 5.13--6.41 & 4.25--6.65  \\ 
1.006--1.738 & 7.64--7.90 & 7.51--7.99  \\ 
2.008--3.107 & 7.39--7.96 & 7.08--8.12  \\ 
3.011--4.475 & 6.71--7.35 & 6.37--7.53  \\ 
\hline
\multicolumn{3}{c}{\em Thermalized CO excitation} \\
0.003--0.369 & 5.89--6.80 & 5.40--7.02  \\ 
1.006--1.738 & 7.59--7.81 & 7.47--7.90  \\ 
2.008--3.107 & 7.03--7.48 & 6.79--7.63  \\ 
3.011--4.475 & 6.70--7.11 & 6.48--7.25  \\ 
\hline
\multicolumn{3}{c}{\em Milky Way--like CO excitation} \\
0.003--0.369 & 5.91--6.82 & 5.43--7.04  \\ 
1.006--1.738 & 7.90--8.12 & 7.79--8.20  \\ 
2.008--3.107 & 7.65--8.09 & 7.41--8.24  \\ 
3.011--4.475 & 7.33--7.81 & 7.08--7.96  \\ 
\hline
\end{tabular}
\end{center}
\end{table}

\begin{table*}
\caption{\rm Example of the line candidates entrying one of the realizations of the CO LFs. (1--2) Sky coordinates of the line candidate; (3) adopted CO--based redshift; (4) signal--to--noise; (5) completeness (see GL19); (6) does the line candidate have a counterpart at optical/NIR wavelengths with matching redshift (see text)? (7) fidelity the line candidate (see GL19); (8) inferred line luminosity; (9) rotational quantum number of the upper energy level of the transition.} \label{tab_catalog}
\begin{center}
\begin{tabular}{ccccccccc}
\hline
  RA  	  & Dec       &$z_{\rm CO}$& S/N & Compl.& c/p? & fid. & $L'$ & $J_{\rm up}$   \\
 $[$deg$]$& $[$deg$]$ &	           &     &	 &      &      &[K\,km\,s$^{-1}$\,pc$^2$]&      \\
 (1)  	  & (2)       & (3)	   & (4) & (5)   & (6)  & (7)  & (8)	          & (9) \\
 \hline
% 53.16063 & -27.77626 & 2.5436     &36.18& 1.00 & Y & 1.00 & $2.69\times10^{10}$ & 3 & 1 \\
% 53.17664 & -27.78551 & 1.3168     &17.50& 1.00 & Y & 1.00 & $8.77\times10^{9}$  & 2 & 1 \\
% 53.17086 & -27.77545 & 2.4534     &15.25& 1.00 & Y & 1.00 & $1.03\times10^{10}$ & 3 & 1 \\
% 53.14350 & -27.78324 & 1.4144     &14.74& 1.00 & Y & 1.00 & $1.78\times10^{10}$ & 2 & 1 \\
% 53.16569 & -27.76991 & 1.5502     &13.98& 1.00 & Y & 1.00 & $1.82\times10^{10}$ & 2 & 1 \\
 53.16063 & -27.77626 & 2.5436     &36.18& 1.00 & Y & 1.00 &  $2.69\times10^{10}$ & 3	\\	
 53.17664 & -27.78551 & 1.3168     &17.50& 1.00 & Y & 1.00 &  $8.77\times10^{9}$  & 2	\\	
 53.17086 & -27.77545 & 2.4534     &15.25& 1.00 & Y & 1.00 &  $1.02\times10^{10}$ & 3	\\	
 53.14350 & -27.78324 & 1.4144     &14.74& 1.00 & Y & 1.00 &  $1.78\times10^{10}$ & 2	\\	
 53.16569 & -27.76991 & 1.5502     &13.98& 1.00 & Y & 1.00 &  $1.82\times10^{10}$ & 2	\\	
 53.16616 & -27.78754 & 1.0952     &11.98& 1.00 & Y & 1.00 &  $6.17\times10^{9}$  & 2	\\	
 53.18138 & -27.77756 & 2.6956     & 9.95& 1.00 & Y & 1.00 &  $2.71\times10^{10}$ & 3	\\	
 53.14822 & -27.77389 & 1.3822     & 9.15& 1.00 & N & 1.00 &  $3.17\times10^{9}$  & 2	\\     
 53.17908 & -27.78062 & 1.0365     & 8.74& 1.00 & Y & 1.00 &  $7.85\times10^{9}$  & 2	\\     
 53.15085 & -27.77440 & 1.3827     & 7.44& 1.00 & N & 1.00 &  $3.64\times10^{9}$  & 2	\\     
 53.16583 & -27.78157 & 1.0964     & 7.31& 1.00 & Y & 1.00 &  $1.97\times10^{9}$  & 2	\\     
 53.14817 & -27.78451 & 3.6013     & 7.12& 0.96 & Y & 1.00 &  $5.59\times10^{9}$  & 4	\\     
 53.14523 & -27.77801 & 1.0985     & 7.10& 0.85 & Y & 1.00 &  $5.08\times10^{9}$  & 2	\\     
 53.15199 & -27.77552 & 1.0963     & 6.78& 1.00 & Y & 1.00 &  $3.56\times10^{9}$  & 2	\\     
 53.16635 & -27.76873 & 1.2942     & 6.11& 1.00 & Y & 0.96 &  $4.57\times10^{9}$  & 2	\\     
 53.17966 & -27.78428 & 0.1129     & 5.96& 1.00 & N & 0.95 &  $1.04\times10^{8}$  & 1	\\     
 53.15192 & -27.78900 & 1.4953     & 5.91& 1.00 & Y & 0.95 &  $5.14\times10^{9}$  & 2	\\     
%53.16731 & -27.77866 & 7.1573     & 5.78& 1.00 & N & 0.82 &  $7.76\times10^{9}$  &	\\     
 53.14544 & -27.78721 & 1.1746     & 5.78& 0.87 & Y & 0.93 &  $3.08\times10^{9}$  & 2	\\     
 53.16513 & -27.76394 & 1.1769     & 5.77& 1.00 & N & 0.75 &  $5.15\times10^{9}$  & 2	\\     
 53.15104 & -27.78691 & 1.7009     & 5.72& 0.95 & Y & 0.87 &  $4.04\times10^{9}$  & 2	\\     
 53.14553 & -27.77757 & 3.6038     & 5.66& 0.92 & Y & 0.61 &  $5.25\times10^{9}$  & 4	\\     
 53.15495 & -27.78709 & 1.0341     & 5.63& 1.00 & N & 0.36 &  $3.39\times10^{9}$  & 2	\\     
 53.14659 & -27.77822 & 1.3821     & 5.56& 0.93 & N & 0.76 &  $3.25\times10^{9}$  & 2	\\     
 53.15702 & -27.78166 & 1.1298     & 5.52& 1.00 & N & 0.72 &  $2.56\times10^{9}$  & 2	\\     
 53.17554 & -27.78809 & 1.3835     & 5.46& 0.46 & N & 0.74 &  $4.05\times10^{9}$  & 2	\\     
%53.16137 & -27.80024 & 4.1614     & 5.43& 1.00 & Y & 0.73 &  $5.57\times10^{9}$  &	\\     
 53.16848 & -27.76772 & 1.2615     & 5.42& 0.92 & Y & 0.49 &  $3.71\times10^{9}$  & 2	\\     
%53.17074 & -27.78950 & 6.1158     & 5.32& 1.00 & N & 0.55 &  $3.94\times10^{9}$  &	\\     
%53.17849 & -27.77839 & 5.2610     & 5.29& 1.00 & Y & 0.69 &  $8.97\times10^{9}$  &	\\     
 53.16946 & -27.79258 & 0.1428     & 5.26& 1.00 & N & 0.55 &  $1.03\times10^{8}$  & 1	\\     
 53.16465 & -27.79427 & 1.0122     & 5.26& 1.00 & N & 0.52 &  $4.60\times10^{9}$  & 2	\\     
 53.14334 & -27.78797 & 3.1259     & 5.25& 1.00 & Y & 0.38 &  $5.31\times10^{9}$  & 4	\\     
 53.14437 & -27.77806 & 0.1873     & 5.23& 1.00 & N & 0.54 &  $2.80\times10^{8}$  & 1	\\     
 53.16572 & -27.79701 & 3.2263     & 5.23& 0.76 & N & 0.55 &  $3.33\times10^{9}$  & 4	\\     
 53.17748 & -27.78064 & 1.1530     & 5.20& 0.96 & Y & 0.53 &  $4.32\times10^{9}$  & 2	\\     
 53.17554 & -27.77674 & 1.2776     & 5.20& 1.00 & N & 0.34 &  $2.14\times10^{9}$  & 2	\\     
 53.14444 & -27.78346 & 2.2128     & 5.12& 1.00 & N & 0.55 &  $5.83\times10^{9}$  & 3	\\     
 53.16161 & -27.77591 & 1.2456     & 5.09& 1.00 & N & 0.24 &  $2.24\times10^{9}$  & 2	\\     
 53.17350 & -27.79211 & 1.3310     & 5.05& 0.93 & N & 0.28 &  $3.61\times10^{9}$  & 2	\\     
 53.14514 & -27.79452 & 1.0398     & 4.97& 1.00 & N & 0.45 &  $4.33\times10^{9}$  & 2	\\     
 53.14967 & -27.78415 & 1.5710     & 4.93& 0.97 & Y & 0.39 &  $3.78\times10^{9}$  & 2	\\     
 53.14690 & -27.78514 & 3.5460     & 4.92& 1.00 & N & 0.50 &  $4.15\times10^{9}$  & 4	\\     
 53.17144 & -27.76966 & 2.2103     & 4.92& 0.96 & N & 0.36 &  $6.23\times10^{9}$  & 3	\\     
 53.14261 & -27.78733 & 1.4269     & 4.90& 1.00 & Y & 0.45 &  $5.03\times10^{9}$  & 2	\\     
\hline
\end{tabular}
\end{center}
\end{table*}

\section{Robustness of the CO Luminosity Functions}\label{sec_robustness}

\begin{figure*}
\begin{center}
\includegraphics[height=0.225\textheight]{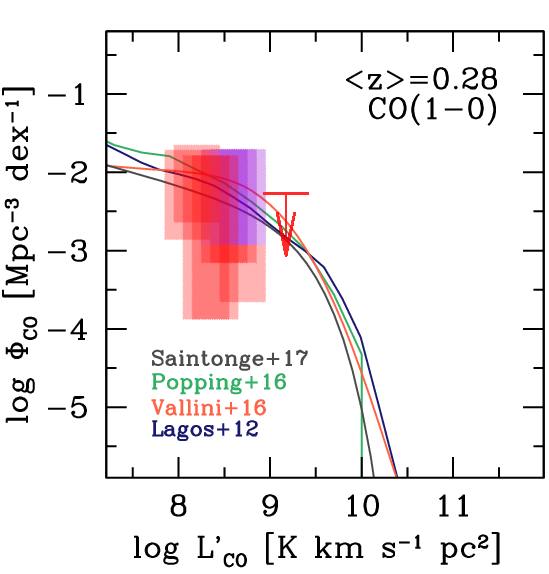}
\includegraphics[height=0.225\textheight]{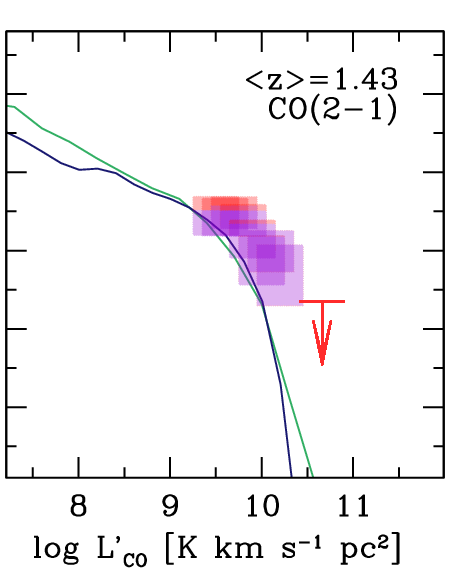}
\includegraphics[height=0.225\textheight]{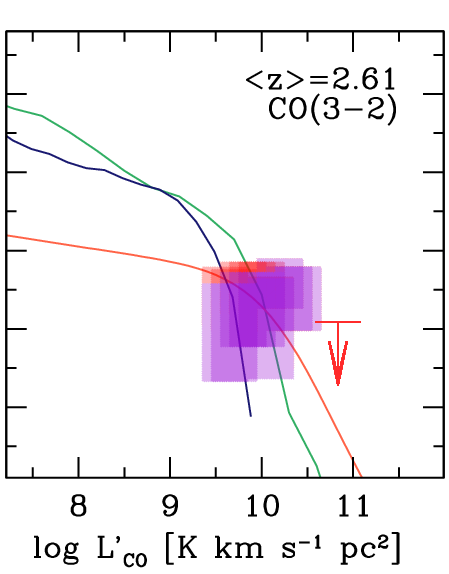}
\includegraphics[height=0.225\textheight]{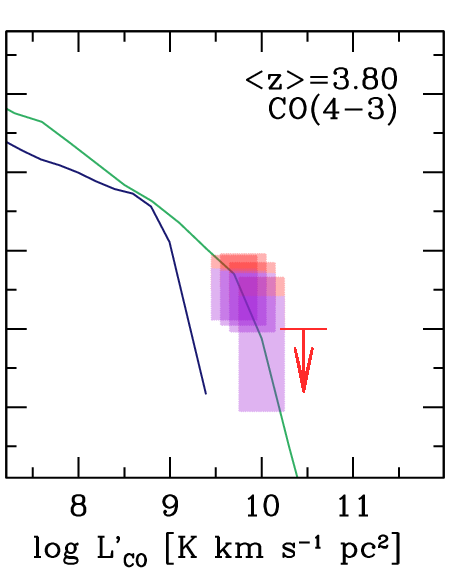}\\
\includegraphics[width=0.55\textwidth]{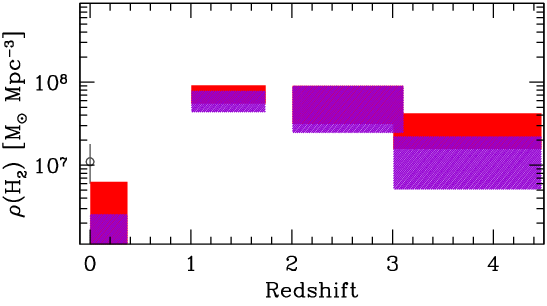}
\end{center}
\caption{CO LFs and evolution of $\rho$(H$_2$) with redshift derived from the entire sample (red shaded boxes) and from the subsample of line candidates that present a counterpart with matching redshifts ($|\delta z|<0.1$). The vertical extent of each box marks the 1-$\sigma$ confidence range. Empirical and semi-analytical model predictions are shown for reference. All of the CO luminosity functions appear consistent in the bright end; at lower luminosities (below $L'_{\rm CO}\approx 10^{10}$ K\,\kms\,pc$^2$) discrepancies arise due to the combined effect of larger rate of false positive candidates at the faint end, and intrinsically fainter counterparts. Our estimates of $\rho$(H$_2$) also appear relatively unchanged if we only focus on sources with matching redshifts, at least up to $z\sim 3$.}
\label{fig_co_lf_secure_z}
\end{figure*}

\begin{figure*}
\begin{center}
\includegraphics[height=0.225\textheight]{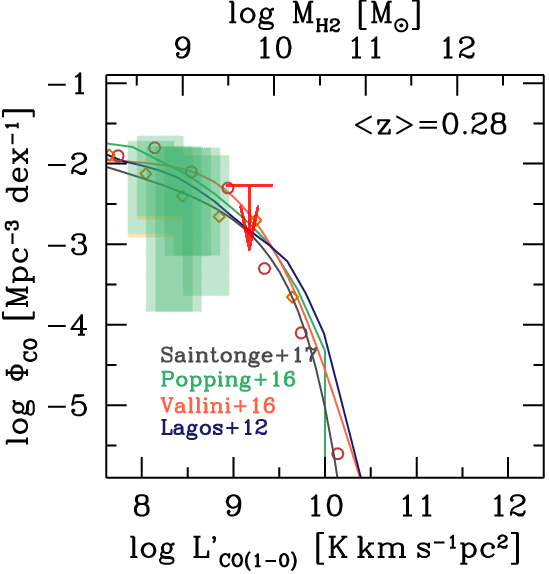}
\includegraphics[height=0.225\textheight]{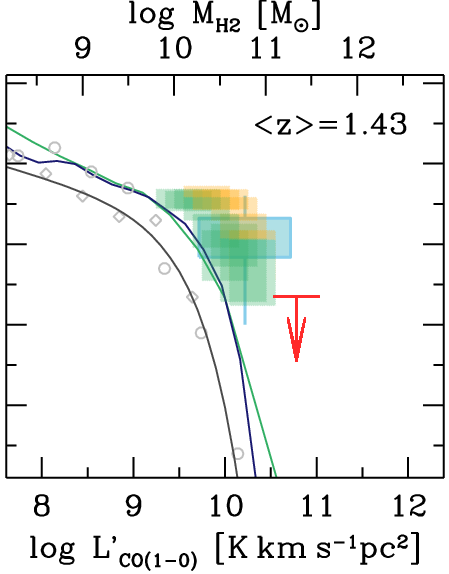}
\includegraphics[height=0.225\textheight]{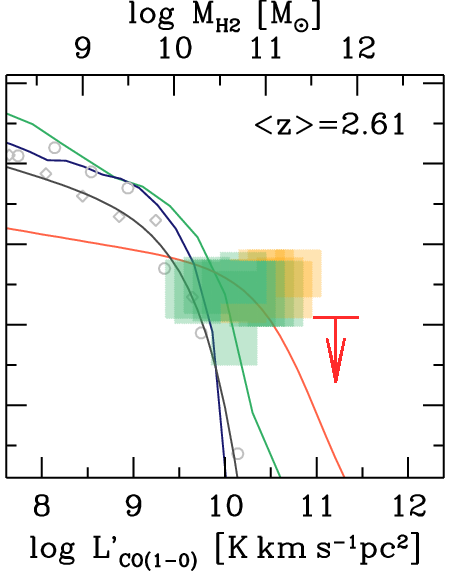}
\includegraphics[height=0.225\textheight]{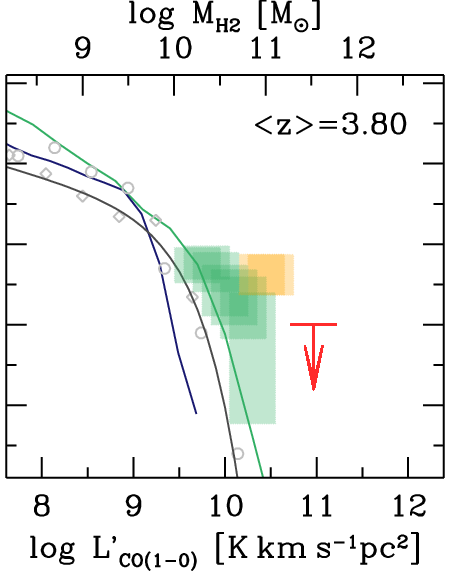}\\
\includegraphics[width=0.55\textwidth]{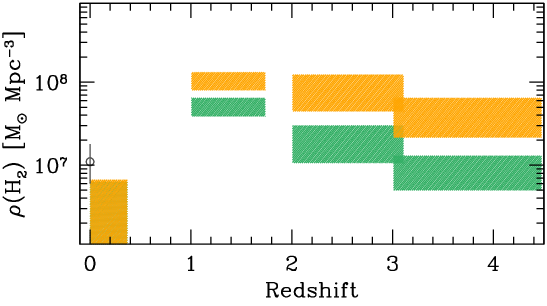}
\end{center}
\caption{CO(1-0) LFs and evolution of $\rho$(H$_2$) with redshift derived assuming two extreme cases of maximal (= thermalized) and minimal (= Milky-Way like) CO excitation, shown in green and orange, respectively. The Milky Way excitation is taken from \citet{weiss07}. The thermalized case assumes $r_{J1}=J^2$ for all the CO transitions. The vertical extent of the boxes marks the 1-$\sigma$ confidence intervals. The \citet{daddi15} CO excitation adopted elsewhere in this paper lies in between these two extreme cases. A high excitation scenario would be in better agreement with the semi-analytical models, especially at the bright end, although it would not be enough to fully account for the discrepancy, especially at $z=1-3$. A high excitation model would also slightly reduce the evolution in $\rho$(H$_2$) between $z=1-3$ and the local universe (by a factor $<2$ with respect to our fiducial assumption), and would naturally predict a faster drop of $\rho$(H$_2$) towards high redshifts. }
\label{fig_co10_lf_excit}
\end{figure*}

Here we test the robustness of the CO LFs constraints from ASPECS LP 3\,mm by creating different realizations of the CO LFs after altering some of the assumptions discussed in the previous section, in particular concerning the fidelity of line candidates, and the CO excitation. The results of these tests are displayed in Figs.~\ref{fig_co_lf_secure_z} and \ref{fig_co10_lf_excit}.

\subsection{Impact of uncertain redshifts / sources with no counterparts}

First, we compare our CO LFs and the constraints on the $\rho$(H$_2$) evolution with redshift against the ones we infer, if we only subselect the galaxies for which a catalog redshift is available, and is consistent with the CO--based redshift within $|\delta z| < 0.1$ (see Fig.~\ref{fig_z_match}). This automatically removes all the line candidates from the line search that lack a counterpart at other wavelengths, as well as potential misassociations with foreground/background galaxies. 

The inferred CO luminosity functions are practically unaltered at their bright end. Small deviations are reported at the faint end, likely due to a combination of two reasons: 1) At the faint end, the impact of false positive candidates is larger. These spurious candidates by definition have counterparts only due to chance alignment, and it is unlikely that such counterparts have matching redshifts. 2) For reasonable ranges of the gas fraction $M_{\rm H2}/M_*$, fainter CO lines are typically associated with fainter stellar emission; these optical/NIR--faint galaxies might have relatively large redshift uncertainties, and might get scattered out of the $|\delta z|<0.1$ selection. 

The direct consequence of these discrepancies is that $\rho$(H$_2$) estimated only using sources with redshift--matching counterparts shows a faster decline at increasing redshifts at $z>3$, compared to our reference estimate, although the two estimates are well within 1-$\sigma$ uncertainties in both the CO LFs and $\rho$(H$_2$) at any redshift. We thus conclude that our results, and in particular the steep evolution in $\rho$(H$_2$) from present day to $z\gsim 1$, are not significantly affected by our treatment of sources without clear counterparts or with ambiguous redshift associations.

\subsection{Impact of CO excitation}

We then examine the impact of the CO excitation assumptions on our estimates of the CO(1-0) LFs and on $\rho$(H$_2$) (the CO LFs of the observed transitions are naturally unaffected by this assumption). We do so by repeating our analysis after assuming two extreme cases: a high excitation case corresponding to thermalized CO up to $J_{\rm up}$=4, and a low excitation scenario where the CO emission is modeled based on the Milky Way disk \citep[see, e.g.,][]{weiss07,carilli13}. A higher (lower) excitation implies fainter (brighter) $L'_{\rm CO(1-0)}$ for a given line observed in a $J_{\rm up}>1$ transition, and therefore lower (higher) values of $M_{\rm H2}$. For reference, our fiducial assumption based on \citet{daddi15} lies roughly half the way between these two extreme cases for the transitions of interest here.

We find that a thermalized CO scenario would mitigate, but not completely solve, the friction between the ASPECS LP 3\,mm CO LFs and the predictions by semi--analytical models. This is further explored in \citet{popping19}. A low--excitation scenario, on the other hand, would exacerbate the tension. Evidence of a strong evolution in $\rho$(H$_2$) between the local universe and $z>1$ is confirmed irrespective of the assumptions on the CO excitation, but for a low--excitation scenario, $\rho$(H$_2$) appears nearly constant at any $z>1$, while it would drop rapidly at increasing redshifts, if a thermalized CO excitation is assumed.

\label{lastpage}

\end{document}